\documentclass[sigconf]{acmart}

\usepackage{booktabs} 
\usepackage{bm}
\usepackage{amsthm}
\usepackage{amssymb}
\usepackage{amsmath}
\usepackage{algorithm}
\usepackage{algorithmic}
\usepackage{hyperref}
\usepackage{epstopdf}
\usepackage{subcaption}
\usepackage{graphicx}
\usepackage{multirow}

\newcommand\numberthis{\addtocounter{equation}{1}\tag{\theequation}}

\renewcommand\footnotetextcopyrightpermission[1]{} 
\acmArticle{4}
\acmPrice{15.00}

\begin{document}
\title{Accelerating E-Commerce Search Engine Ranking by Contextual Factor Selection}

\author{Yusen Zhan}
\affiliation{%
  \institution{Alibaba Group}
  \streetaddress{P.O. Box 1212}
  \city{Hangzhou}
  \state{China}
  \postcode{43017-6221}
}
\email{yusen.zys@alibaba-inc.com}

\author{Qing Da}
\affiliation{%
	\institution{Alibaba Group}
	\streetaddress{P.O. Box 1212}
	\city{Hangzhou}
	\state{China}
	\postcode{43017-6221}
}
\email{daqing.dq@alibaba-inc.com}

\author{Fei Xiao}
\affiliation{%
	\institution{Alibaba Group}
	\streetaddress{P.O. Box 1212}
	\city{Hangzhou}
	\state{China}
	\postcode{43017-6221}
}
\email{guren.xf@alibaba-inc.com}

\author{An-xiang Zeng}
\affiliation{%
	\institution{Alibaba Group}
	\streetaddress{P.O. Box 1212}
	\city{Hangzhou}
	\state{China}
	\postcode{43017-6221}
}
\email{renzhong@taobao.com}
\author{Yang Yu}
\affiliation{%
 \institution{National Key Laboratory for Novel Software Technology,\\ Nanjing University}
 \city{Nanjing}
 \country{China}
 \postcode{210023}
}
\email{yuy@lamda.nju.edu.cn}

\renewcommand{\shortauthors}{}

\begin{abstract}
In industrial large-scale search systems, such as Taobao.com search for commodities, the quality of the ranking result is getting continually improved by introducing more factors from complex procedures, e.g., deep neural networks for extracting image factors. Meanwhile, the increasing of the factors demands more computation resource and raises the system response latency. It has been observed that a search instance usually requires only a small set of effective factors, instead of all factors. Therefore, removing ineffective factors significantly improves the system efficiency. This paper studies the \emph{Contextual Factor Selection} (CFS), which selects only a subset of effective factors for every search instance, for a well balance between the search quality and the response latency. We inject CFS into the search engine ranking score to accelerate the engine, considering both ranking effectiveness and efficiency. The learning of the CFS model involves a combinatorial optimization, which is transformed as a sequential decision-making problem. Solving the problem by reinforcement learning, we propose the RankCFS, which has been assessed in an off-line environment as well as a real-world on-line environment (Taobao.com). The empirical results show that, the proposed CFS approach outperforms several existing supervised/unsupervised methods for feature selection in the off-line environment, and also achieves significant real-world performance improvement, in term of service latency,  in daily test as well as Singles' Day Shopping Festival in $2017$.
\end{abstract}

%
%


\maketitle

\section{Introduction}
Information retrieval and machine learning applications play an important role in industrial and commercial scenarios, ranging from web searching engine (Google.com, Baidu.com, etc.) to e-commence websites (Taobao.com, Amazon.com). The major applications, ie., search and recommendation, usually require to rank a large set of data items in terms of response to users' requests under an on-line circumstance. To support these applications, there are generally two issues: a) Effectiveness such as how accurate and reliable the search results in the final ranking list are  and b) Efficiency as how fast the search engine's response to the user's queries in a timely manner and whether the computational burden of ranking is as low as possible from a system's perspective. It is a challenge to address both issues in large-scale applications for providing excellent user experience and an efficient performance solution.

Generally speaking, to address the high computational cost of more deep models as well as large-scale traffic requests, the search engine system has to degrade the service level in the aspect of effectiveness, i.e., reducing the number of recalled items, off-lining some unnecessary service and so on, in oder to avoid access delay or even unavailability, which severely affects the users' experience. Though successful, these methods only adopts a compromise between the search engine processing performance and the service availability in a hard way, which means there must be unnecessary sacrifice of revenue in real world business practice. Consequently, this raises a question whether we are able to design a "soft" or "intelligent" solution, which allows to achieve both of effectiveness and efficiency.  

\begin{figure}[t]
	\centering
	\includegraphics[width=0.6\linewidth]{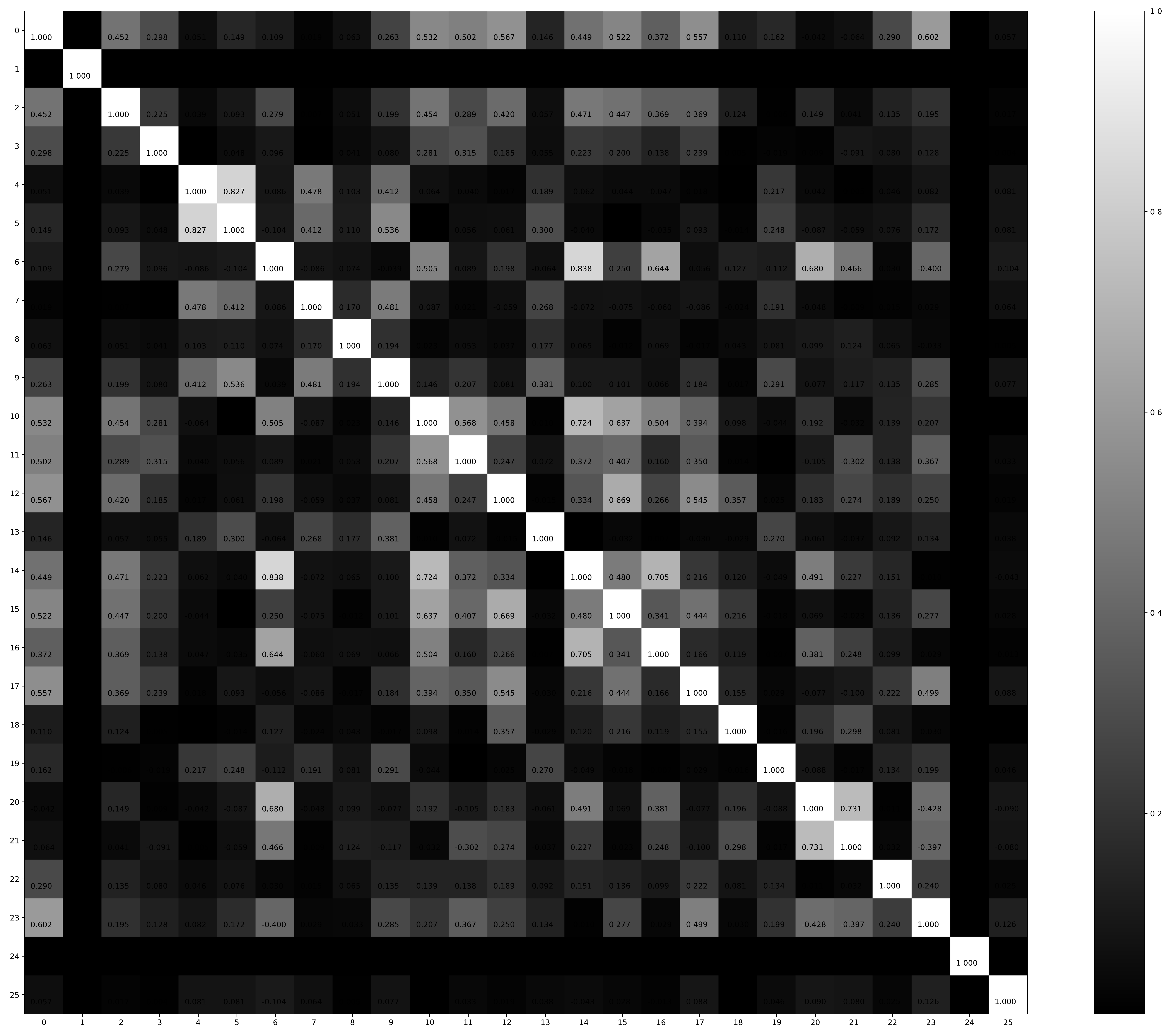}
	\caption{The element-wise Pearson product-moment correlation coefficients between selected factors. The darker the block is,  the less correlated the corresponding factors are, and vice versa}
	\label{fig:scoreforanalysispay}
\end{figure}

The answer seems to be promising. Liu \textit{et al.} proposed a cascading ranking model to address the trade-off between effectiveness and efficiency in the large-scale e-commerce search applications \cite{Liu:2017:CRO:3097983.3098011}. Their method mainly focuses on reducing the number of items in the ranking process, however, their model sheds a light on us to optimize the e-commerce search engine in another possible way. In a search engine, a set of factors is applied to the ranking process and we conjecture that not all of those factors are necessary in the real-world applications. After thorough investigation in the real world operational environment, we discover that there are still relatively high correlations between those ranking factors in our system. 
See Figure \ref{fig:scoreforanalysispay} for details. 
Therefore, on the one hand, there exists redundancy of factors in our on-line operational environment. On the other hand, we also realize that the conversion rates vary on items under different contexts\footnote{Here, the $\langle u, q \rangle$ user-query pair denotes the context.}. For instance, the users with higher purchase power always have a higher conversion rate under some long-tail (low-frequency) queries. Based on aforementioned analysis, we consider that some computational efficient factors may be sufficient for achieving effectiveness under such contexts. The above  observations shows the possibility of keeping the effectiveness by carefully selecting a subset of all factors under certain circumstances, which indeed is a standard combinatorial optimization problem, but with auxiliary context description.

Combinatorial optimization is a fundamental problem in computer science. Recently, Bello \textit{et al.} show that reinforcement learning is capable of solving combinatorial optimization problem like TSP via pointer network \cite{vinyals2015pointer,bello2016neural}. In this paper,  we try to address above challenges by designing an innovative model via reinforcement learning algorithms. We formally define our optimization problem with a general framework and a loss function which is able to reflect both of ranking effectiveness and efficiency. Then, we transform the contextual combinatorial optimization problem to a sequential decision-making one by incorporating the  contextual setting and factor selection into the state and action of an MDP, respectively. The reward is designed to encourage to save the computational cost of factors as well as ensure that the ranking results are still effective. The final solution can be obtained via the state-of-the-art reinforcement learning algorithms such as Asynchronous Advantage Actor-Critic (A3C) in this paper. Based on the correlation among factors as well as the context dependency in our system, our method is capable of handling contextual factor selection in terms of user and query, while ensuring to minimized the influence on business indicators, i.e., gross merchandise value (GMV), click-through rate (CTR) and so on. 
 
We show our algorithm outperforms comparative algorithms in both of off-line and on-line evaluation. In Singles' Day Shopping Festival, $2017$, we also demonstrate the capabilities of this new method in real-world large-scale system. 

The contributions of this paper can be summarized as: \textit{i)} injecting contextual factor selection into search engine ranking score for engine acceleration, \textit{ii)} formulating the contextual factor selection for ranking as a contextual combinatorial optimization problem, \textit{iii)} deriving a reinforcement learning based solution to the proposed optimization problem, \textit{iv)} demonstrating the effectiveness of our technique in both of off-line and on-line environments.

The rest of the paper is organized as follows: Section 2 introduces the background; Section 3 provides some related work; In section 4, we define our problem in a view of optimization; Section 5 proposes actor-critic method to resolve the problem in section 4; We show the  experimental results in section 7. Finally, section 8 summarizes the whole paper. 


\section{Ranking in E-Commerce Search}
\label{subsection:ltrbackground}
Suppose that $\mathcal{O}$ is the set of all available items in the database, $\mathcal{Q}$ is the set of all possible queries and $\mathcal{U}$ denotes the set of all users' information. Let $\left\{\langle u,q \rangle_1,\langle u,q \rangle_2,\dots,\langle u,q \rangle_m\right\}$ be a set of user-query pairs, where $\langle u,q \rangle_i \in \mathcal{U}\times\mathcal{Q}$ denotes the $i$-th user-query pair from the search requests. $O_i=\{o_{i,1},o_{i,2},\cdots,o_{i,n_i}\}$ is the set of items associated with the  $i$-th user-query request, where, $n_i$ is the number of the related items. The ranking problem in e-commerce scenario can be then formally defined as a task to generate a permutation function $\sigma_i \in \Sigma_i$, where $\sigma_i$is an one-to-one correspondence from $\{1,2,\dots,n_i\}$ to itself and $\Sigma_i$ denotes the set of all the possible permutations on $\mathcal{O}_i$. The goal is to maximize the probability of purchase under the permutation. The permutation is usually generated by a ranking function $F(\langle u,q \rangle_i,o_{i,j}) \to \mathbb{R}$ which scores each item $o_{i,j} \in O_i$ for the request $\langle u,q \rangle_i$.
 Let $\bm{x}^{i,j} \in \mathbb{R}^n$ be the corresponding factor vector of item $o_{i,j} \in \mathcal{O}_i $ under query $\langle u,q \rangle_i$, where $i=1,2,\dots,m$; $j=1,2,\dots,n_i$. Some factors in the factor vector depends on the user-query pair $\langle u,q \rangle_i$ and the item $a_{i,j}$. Without loss of generality,  the ranking model is defined by
\begin{equation}
F(\langle u,q \rangle_i,o_{i,j})=f\left(\bm{x}^{i,j}\right),
\label{eq:rankingmodel}
\end{equation}
where $f: \mathbb{R}^n \to \mathbb{R}$ is the ranking function. It could be any function such as a linear model, a deep neural network or a tree model. 

The ranking function is usually trained from a dataset $\mathcal{D}=\left\{\left(\langle u,q \rangle_i,\bm{x}^{i}, \bm{y}_i\right)\right\}_{i=1}^N$ logged from the real system, where $N$ is the number of training examples and $\bm{y}_i=\{y_{1,i},y_{2,i},\dots,y_{i,n_i}\}$ denotes the labels associates with items. Specifically, $y_{i,j} \in \mathcal{Y}=\{\text{view},\text{click},\text{buy}\}$ represents the feedback of the user on the $j$-th item. The training can be conducted in point-wise way~\cite{Cooper1992Probabilistic,learning-to-rank-using-classification-and-gradient-boosting}, pair-wise way~\cite{Freund2003An,Burges2005Learning,Zheng2007A}, or list-wise way~\cite{Xu2007AdaRank,learning-to-rank-from-pairwise-approach-to-listwise-approach,Burges2010From}. It is worth noting that, in this paper we assume that a trained ranking function is given and consider the general case that the ranking function is provided as a black box, i.e., with no access to the gradient or even the Hessian matrix.

\section{Related Work}
There are a lot of work that attempts to resolve the effectiveness and efficiency challenge and we review will some of them.

Cascade learning is originally proposed to address the effectiveness and efficiency issue in traditional classification and detection problems such as fast visual object detection ~\cite{bourdev2005robust,schneiderman2004feature,Viola2003Rapid}. Liu \textit{et al.} develop a cascade ranking model for a large-scale e-commerce search system and deploy it in Taobao.com~\cite{Liu:2017:CRO:3097983.3098011}. However, they only exploit optimization in terms of the number of ranking items, we mainly focuses on the factor usage during the ranking process. 

Feature selection tries to remove irrelevant and/or redundant features to improve learning performance \cite{guyon2003introduction}. Traditional feature selection techniques roughly fall into two categories, i.e., filter methods and wrapper methods. Filter methods use learner-irrelevant measurements to evaluate and select features, such as information gain and Relief~\cite{kira1992feature}. Wrapper methods involve the final learner in the feature selection process, such as using the accuracy as the evaluation criterion for the goodness of features. Liu \textit{et al.} proposed  the TEFE (Time-Efficient Feature Extraction) approach,
which balances between the test accuracy and test time cost by extracting a proper subset of features for each test object \cite{liu2008tefe}. In learning to rank literature, Feature selection is a common strategy to improve the efficiency. In generally, a set of crucial factors are selected from a complete set of all possible factors according to some criteria such as importance to the ranking \cite{Geng:2007:FSR:1277741.1277811,Wang:2010:LER:1835449.1835475,Wang:2010:RUT:1871437.1871452}.
Geng \textit{et al.} propose a selection method based on factor importance in a query-free manner, but they do not consider the real computational cost and query-dependent factor~\cite{Geng:2007:FSR:1277741.1277811}. There are also some methods are query-dependent, in which the cost (delay) of the query is considered~\cite{Wang:2010:LER:1835449.1835475,Wang:2010:RUT:1871437.1871452}. In contrast, we consider the computational cost (delay) of individual factor.

Ensemble pruning is a class of approaches that tries to select a subset of learners (factors) to comprise the ensemble learner \cite{zhou2012ensemble}. Recently, Benbouzid \textit{et al.} apply Q-learning algorithm to ensemble pruning, in which a reinforcement learning agent tries to decide whether or not to skip the base learner. However, their method is context-free and lacks of evaluation in a real-world large-scale application.

\section{Contextual Factor Selection for Ranking}
\subsection{The CFS framework}
\label{subsection:efficiency}
In this subsection, we describes a general framework of \emph{Contextual Factor Selection} (CFS), for constructing a search engine optimizer which achieves both of effectiveness and efficiency in terms of e-commerce search engine. As mentioned above, a factor vector $\bm{x}_{i,j} \in \mathbb{R}^p$ is assigned to the corresponding item $o_{i,j} \in \mathcal{O}_i$, in which each dimension of the factor vector is calculated on-line and varies on the computational cost.  Let $\bm{x}^{i,j}=\{x^{i,j}_1,x^{i,j}_2,\dots,x^{i,j}_p\}$ be the factor vector associated with a cost vector $\bm{c}=\{c_1,c_2,\dots,c_p\}$, where $c_k$ denotes the computational cost of the $k$-th factor. Let $\Omega$ be the set of all factors and $S$ be a subset of $\Omega$. The indicator function of a subset $S$ of the set $\Omega$ defined as
\begin{align}
\mathbb{I}_S(k)= \begin{cases}
1 & x_k \in S,\\
0 & x_k \notin S.
\end{cases}
\end{align}
From a practical point of view, some of factors are not necessary in terms of ranking. For example, given a set of factors $\Omega=\{x^{i,j}_k \mid k=1,2,\dots,p\}$, a subset $S$ of $\Omega$ with highly confident factors might be sufficient under some contexts. Therefore, given an item $o_{i,j}$ and indicator function $\mathbb{I}_S$, the computational cost function can be written as $\sum_{k=1}^p \mathbb{I}_S(k)c_k$, where the indicator function determines whether or not we use the factor to participate the sorting process\footnote{We mainly consider the computational cost of the factors while ignoring other costs.}. Thus, given a set of items $\mathcal{O}_i$, the total computational cost  is 
\begin{align}
\sum_{j=1}^{n_i}\sum_{k=1}^p \mathbb{I}_S(k) c_k,
\end{align}

As defined in Equation \ref{eq:rankingmodel}, the ranking model with all factors can be written as $F_{\Omega}(o_{i,j})=f(x^{i,j}_1,x^{i,j}_2,\dots,x^{i,j}_p)$ and the one with a subset $S$ is written as 
\begin{align}
F_{S}(o_{i,j})=f\left(\mathbb{I}_S(1)x^{i,j}_1,\mathbb{I}_S(2)x^{i,j}_2,\dots,\mathbb{I}_S(p)x^{i,j}_p\right)
\end{align}
Intuitively, we can treat the permutation generated by $F_{\Omega}$ as the optimal one since it includes all the factors we have during the ranking process. Thus, given a $\langle u,q\rangle_i$ request, the objective is
\begin{equation}
\min\limits_{S\subseteq \Omega} \;D^{\mathcal{O}_i} (F_{\Omega} || F_S) + \lambda {n_i} \sum_{k=1}^p \mathbb{I}_S(k) c_k,
\label{eq:generalobjective}
\end{equation}
where $D^{\mathcal{O}_i}\left(F_{\Omega} || F_S \right)$ denotes the distance between function $F_{\Omega}$ and $F_S$ over the item set ${\mathcal{O}_i}$, which could be any distance between two functions, i.e., Kullback-Leibler divergence~\cite{kullback1951information}, the second term is the computational costs of factors in the set $S$, $\lambda>0$ is the trade-off parameter and $n_i$ is the number of items in query $i$. Intuitively, the objective implies that it reduces the usage of factors as many as possible, while approximating the original ranking function $F_{\Omega}$ by function $F_S$ as close as possible.  

However, Equation \ref{eq:generalobjective} is intractable even for a single $\langle u,q\rangle_i$ request, which is able to reduced to the optimal subset selection problem. Consequently, it is a NP-hard problem in general \cite{davis1997adaptive,natarajan1995sparse}. Moreover, we need to do the contextual factor selection, i.e., solving a general NP-hard problem for every $\langle u,q\rangle_i$, which is impractical in a large-scale system even with a small number of contexts. To overcome this challenge, we try to generalize the solution of Equation \ref{eq:generalobjective} at the contextual level. That is, we do not directly search the optimal subset $S^{\star}$ and define :
\begin{align}
S_{\langle u,q\rangle}=H\left(\langle u,q \rangle \mid \bm{\theta}\right)
\end{align}
where $H$ is a model parameterized by $\bm{\theta}$ and the user-query pair $\langle u,q\rangle$ characterizes the context. Such formulation reduces the solution space to a global parameter from the original multiple optimal subset selection problems, based on the assumption that similar $\langle u,q\rangle$ representations should have similar optimal subset structure. Thus, our goal is to search for the global parameter vector $\bm{\theta}$ to minimize the loss defined in Equation \ref{eq:generalobjective} over all the $\langle u,q\rangle$ requests. 

To illustrate our method, we adopt the linear ranking functions as a demonstration, and other representations, i.e., deep neural network and tree based ranking function, can be derived by similar way. In the linear setting, the score of item $o_{i,j}$ under user-query $\langle u,q\rangle_i$ is 
\begin{align}
f(x^{i,j}_1,x^{i,j}_2,\dots,x^{i,j}_p)=\sum_{k=1}^p w_k^{i} x^{i,j}_k,
\end{align}
where $w_k^{i}$ is the corresponding weight of factor $x^{i,j}_k$.

\begin{figure}[t]
	\centering
	\includegraphics[width=1\linewidth]{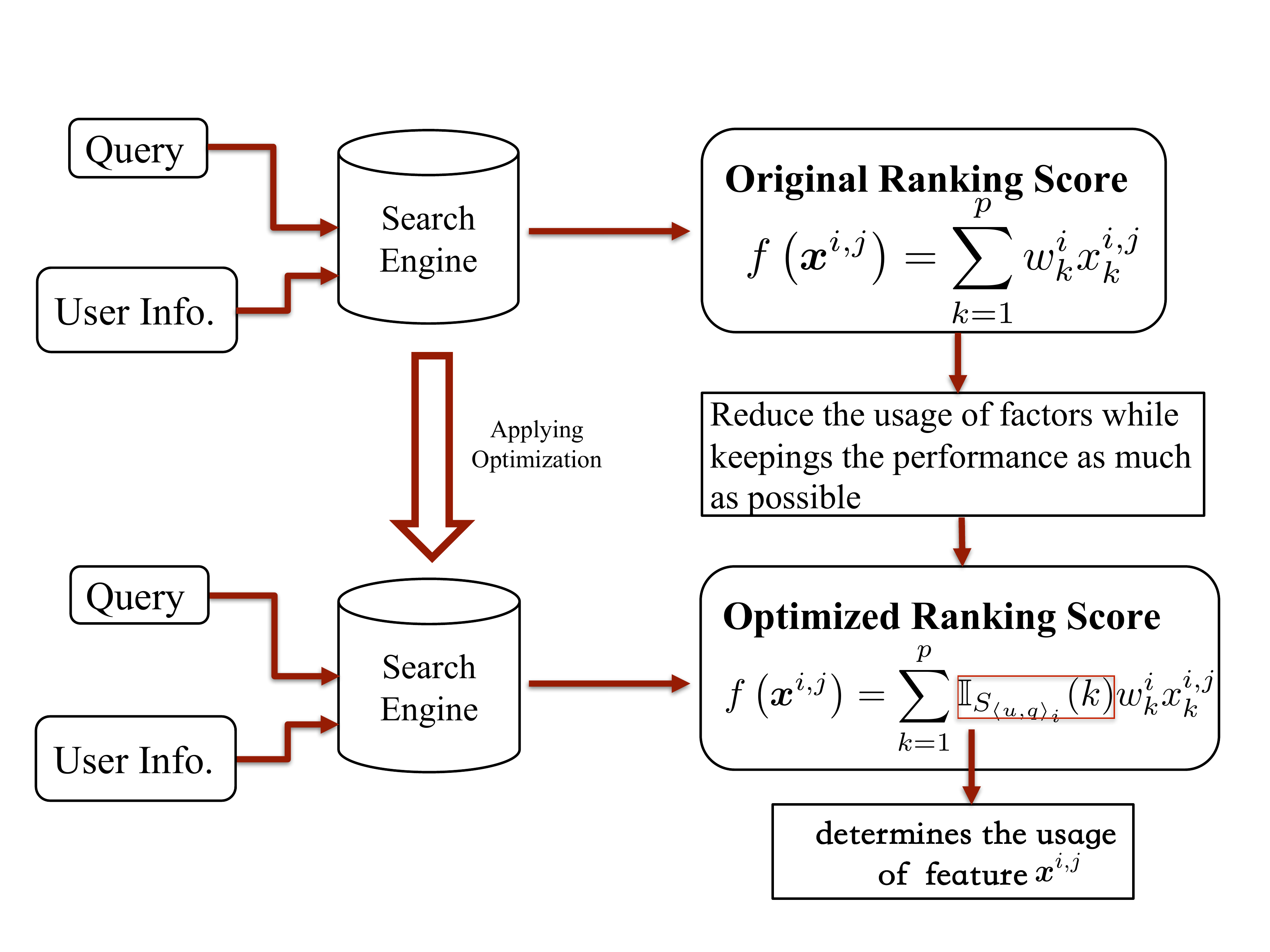}
	\caption{Ranking optimization illustration.}
	\label{fig:rlforrankingoptimization}
	\vspace{-5pt}
\end{figure}

In another point of view, the permutation $\sigma_i \in \Sigma_i$ significantly depends on the factors are used to calculate the scores. Formally, given an user-query pair $\langle u,q\rangle_i$ and a corresponding weight vector $\bm{w}^{\langle u,q\rangle_i}$, the linear ranking function 
\begin{align}
f\bigg(\mathbb{I}_{S_{\langle u,q\rangle_i}}&(1)x^{i,j}_1,\mathbb{I}_{S_{\langle u,q\rangle_i}}(2)x^{i,j}_2,\dots,\mathbb{I}_{S_{\langle u,q\rangle_i}}(n)x^{i,j}_p \bigg) \nonumber \\ 
& \quad=\sum_{k=1}^p \mathbb{I}_{S_{\langle u,q\rangle_i}}(k) w_k^{i} x^{i,j}_k
\end{align}
where $\mathbb{I}_{S_{\langle u,q\rangle_i}}(k)$ is the indicator function, which depends on the user-query pair $\langle u,q\rangle_i$. For convenience, $\mathbb{I}_{S_{\langle u,q\rangle_i}} \in \{0,1\}^p$ denotes the binary vector with respect to the factor vector $\bm{x}^{i,j}$. Therefore, the ranking permutation $\sigma_i$ highly depends on the ranking function $f(\cdot)$ and the indicator function $\mathbb{I}_{S_{\langle u,q\rangle_i}}$, assuming the weight vector is fixed if the ranking model is given. 
Thus, the crucial part of ranking optimization is to learn an indicator function $\mathbb{I}_{S_{\langle u,q\rangle_i}}$ to determine the utilization of the factors. See Figure \ref{fig:rlforrankingoptimization} for illustration. To simplify the notation, we write $\mathbb{I}_{S_{\langle u,q\rangle_i}}$ as $\mathbb{I}_{\bm{\theta}}$, where the parameter $\bm{\theta}$ characterizes the factor subset $S_{\langle u,q\rangle_i}$.  Thence, the ranking permutation $\sigma_i^{\mathbb{I}_{\bm{\theta}}}$ is induced by the ranking function $f(\cdot)$ and the indicator function $\mathbb{I}_{\bm{\theta}}$. Thus, we can rewrite the distance function $D^{\mathcal{O}_i}\left(F_{\Omega} || F_S \right)$ as
$D^{\mathcal{O}_i}\left(\sigma_{\Omega} || \sigma_S \right)$, where $\sigma_{\Omega}$ and $\sigma_S$ are permutations reduced by ranking function $F_{\Omega}$ and $F_S$, respectively.

\subsection{CFS with Pairwise Ranking Loss}
With the optimal ranking permutation $\sigma_{\Omega}$ above, then we define the distance over a item set $\mathcal{O}_i$ between a permutation $\sigma_i^{\mathbb{I}_{\bm{\theta}}}$ and the optimal ranking permutation $\sigma_{\Omega}$ as 
\begin{align}
\label{eq:avaragepairwiseloss}
D^{\mathcal{O}_i}(\sigma_{\Omega}||\sigma_i^{\mathbb{I}_{\bm{\theta}}})= \frac{2}{n_i(n_i-1)}\sum_{    \substack{  j,k=1,j\neq k \\ \sigma_{\Omega}(j)\ge \sigma_{\Omega}(k)  }    }^{n_i}\mathbf{1}(\sigma_i^{\mathbb{I}_{\bm{\theta}}}(j)<\sigma_{i}^{\mathbb{I}_{\bm{\theta}}}(k)),
\end{align}
where $\mathbf{1}(\sigma_i^{\mathbb{I}_{\bm{\theta}}}(j)<\sigma_{i}^{\mathbb{I}_{\bm{\theta}}}(k))$ equals $1$ if $\sigma_i^{\mathbb{I}_{\bm{\theta}}}(j)<\sigma_{i}^{\mathbb{I}_{\bm{\theta}}}(k)$ and $0$ otherwise. The definition of the distance $D$ is the analogue of the averaged pairwise loss in learning to rank literature. The distance measures that how \emph{far away} is the induced permutation to the optimal one in terms of ranking pairs. 

With the distance and total cost function defined above, our goal is, given a user-query pair $\langle u,q\rangle_i$, the corresponding item set $\mathcal{O}_i$ and the ranking function $f$, to learn an indicator function $\mathbb{I}_{\bm{\theta}}$ such that minimizes the the distance $D$ function and the total computational costs. Formally, the objective in Equation \ref{eq:generalobjective} can be further rewritten as 
\allowdisplaybreaks
\begin{align}
\mathcal{L}(\langle u,q \rangle_i, &\mathcal{O}_i, f  \mid \bm{\theta})= D^{\mathcal{O}_i}(\sigma_{\Omega}||\sigma_i^{\mathbb{I}_{\bm{\theta}}}) +  \lambda \sum_{j=1}^{n_i} \sum_{k=1}^p \mathbb{I}_{\bm{\theta}}( k) c_k \nonumber \\
&= \underbrace{   \frac{2}{n_i(n_i-1)} \sum_{  \substack{  j,k=1,j\neq k \\ \sigma_{\Omega}(j)\ge \sigma_{\Omega}(k)  }    }^{n_i}\mathbf{1}(\sigma_i^{\mathbb{I}_{\bm{\theta}}}(j)<\sigma_{i}^{\mathbb{I}_{\bm{\theta}}}(k))   }_{\text{Ranking Effectiveness}} \nonumber \\
& \quad +\lambda\underbrace{ {n_i}\sum_{k=1}^p \mathbb{I}_{\bm{\theta}}(k)  c_k}_{\text{Ranking Efficiency}}
\label{Eq:linearobjective}\numberthis 
\end{align}

\section{RankCFS: A Reinforcement Learning Approach}
As mentioned in section \ref{subsection:efficiency}, the optimization problem defined in Equation \ref{Eq:linearobjective} is NP-hard in general case and finding the exact solution is computationally intractable. Inspired by recent work \cite{bello2016neural,busa2012fast}, we propose to optimize the factor usage using reinforcement learning framework in order to learn an indicator function $\bm{\theta}$, by transforming the assignment of each element in the indicator vector as a sequential decision-making problem. We call it RankCFS.

\subsection{Reinforcement Learning and Actor-Critic Methods}
In this subsection, we will review some basic concepts in reinforcement learning. This subsection could be skipped If the readers are similar with reinforcement learning. 

In reinforcement learning, an agent must sequentially select actions to maximize its total expected pay-off. These problems are typically formalized as Markov decision processes (MDPs) with a tuple of $\left \langle \mathcal{S}, \mathcal{A}, \mathcal{P}, \mathcal{R}, \gamma \right \rangle$, where $\mathcal{S} \subseteq \mathbb{R}^{d}$ and $\mathcal{A} \subseteq \mathbb{R}^{m}$ denote the state and action spaces. $\mathcal{P}: \mathcal{S} \times \mathcal{A} \times \mathcal{S} \rightarrow [0,1]$ represents the transition probability governing the dynamics of the system, $\mathcal{R}: \mathcal{S} \times \mathcal{A} \rightarrow \mathbb{R}$ is the reward function quantifying the performance of the agent and $\gamma \in (0,1)$ is a discount factor specifying the degree to which rewards are discounted over time. At each step $t$, the agent is in state $\bm{s}_{t} \in \mathcal{S}$ and must choose an action $\bm{a}_{t} \in \mathcal{A}$, transitioning it to a successor state $\bm{s}_{t+1} \sim p(\bm{s}_{t+1}|\bm{s}_{t}, \bm{a}_{t})$ as given by $\mathcal{P}$ and yielding a reward $r_{t}$. A policy $\pi: \mathcal{S} \times \mathcal{A} \rightarrow [0,1]$ is defined as a probability distribution over state-action pairs, where $\pi\left(\bm{a}_{t}|\bm{s}_{t}\right)$ denotes the probability of choosing action $\bm{a}_{t}$ at state $\bm{s}_{t}$. 

\emph{Policy gradients}~\cite{kober2009policy,Sutton}  are a class of reinforcement learning algorithms that have shown successes in solving complex robotic problems~\cite{kober2009policy}. Such methods represent the policy $\pi_{\bm{\theta}}(\bm{a}_{t}|\bm{s}_{t})$ by an unknown vector of parameters $\bm{\theta} \in \mathbb{R}^{d}$. The goal is to determine the optimal parameter vector $\bm{\theta}^{\star}$ that maximize the expected discounted cumulative reward: 
\begin{align}
\mathcal{J}(\bm{\theta}) = \sum_{\bm{\tau}} P(\bm{\tau}|{\bm{\theta}}) \mathfrak{R}(\bm{\tau}),
\end{align}
where $\bm{\tau} = [\bm{s}_{0:T}, \bm{a}_{0:T}]$ denotes a trajectory over a possibly finite horizon $T$. The probability of acquiring a trajectory, $ P(\bm{\tau}|{\bm{\theta}})$, under the policy parameterization $\pi_{\bm{\theta}}(\cdot)$ and discounted cumulative reward $\mathfrak{R}(\bm{\tau})$ is given by: 
\begin{align}
P(\bm{\tau}|{\bm{\theta}}) &= p_{0}(\bm{s}_{0}) \prod_{t=0}^{T-1} p\left(\bm{s}_{t+1}\mid\bm{s}_{t}, \bm{a}_{t}\right)\pi_{\bm{\theta}} (\bm{a}_{t}\mid\bm{s}_{t}), \\
\mathfrak{R}(\bm{\tau}) &=  \sum_{t=0}^{T} \gamma^t r_{t+1},
\end{align}
with an initial state distribution $p_{0} : \mathcal{X} \rightarrow [0,1]$. 
Policy gradient methods, such as episodic REINFORCE~\cite{williams1992simple} and Natural Actor Critic~\cite{bhatnagar2009natural,peters2008natural}, typically employ a lower-bound on the expected return $\mathcal{J}(\bm{\theta})$ for fitting the unknown policy parameters $\bm{\theta}$. To achieve this, such algorithms generate trajectories using the current policy $\pi_{\bm{\theta}}$, and then compare performance with a new parameterization $\tilde{\bm{\theta}}$. As detailed in~\cite{kober2009policy}, the policy gradient of $\mathcal{J}(\bm{\theta})$ can be estimated using the the likelihood ratio trick as 
\begin{align}
\label{eq:objlowerbound} 
\nabla_{\bm{\theta}} \mathcal{J}(\bm{\theta})=\sum_{\bm{\tau}} P(\bm{\tau}|{\bm{\theta}}) \nabla_{\bm{\theta}} \log{P(\bm{\tau}|{\bm{\theta}})} \mathfrak{R}(\bm{\tau})
\end{align}
which is usually approximated with empirical estimate for $m$ sample trajectories under the policy $\pi_{\bm{\theta}}$, i.e., $\frac{1}{m}\sum_{j=1}^m  \nabla_{\bm{\theta}} \log{P(\bm{\tau^j}|{\bm{\theta}})} \mathfrak{R}(\bm{\tau^j})$. The gradient can be applied in every step $t$ and further improved by introducing a learned bias $V^{\pi_{\bm{\theta}}}(s_t|\mu)$ to reduce the variance of this estimate as in~\cite{mnih2016asynchronous}
\begin{align}
d\theta \leftarrow \nabla_{\bm{\theta}} \log{\pi_{\bm{\theta}}(a_t|s_t)}(\mathfrak{R_t} - V^{\pi_{\bm{\theta}}}(s_t|\mu))
\end{align}
where $\mathfrak{R_t}=\sum_{i=t}^{T} \gamma^{i-t} r_{i}$ is the discounted cumulative reward from step $t$ and $V^{\pi_{\bm{\theta}}}(s_t|\mu)$ is the function approximation of $\mathfrak{R_t}$ parameterized by $\mu$, of which the gradient is
\begin{align}
d\mu \leftarrow (\mathfrak{R_t}-V^{\pi_{\bm{\theta}}}(s_t|\mu)) \nabla_{\mu} V^{\pi_{\bm{\theta}}}(s_t|\mu)
\end{align}

\subsection{Converting CFS to MDP Setting}
It is possible to learn a factor subset, in which a subset of factors are chosen for the ranking process, by a reinforcement learning policy, instead of approximating the indicator vector $\mathbb{I}_{\bm{\theta}}$ directly. However, it results in a combinatorial action space which leads to computational intractability and searching failure with a high probability.  

\begin{figure}[t]
	\centering
	\includegraphics[width=1\linewidth]{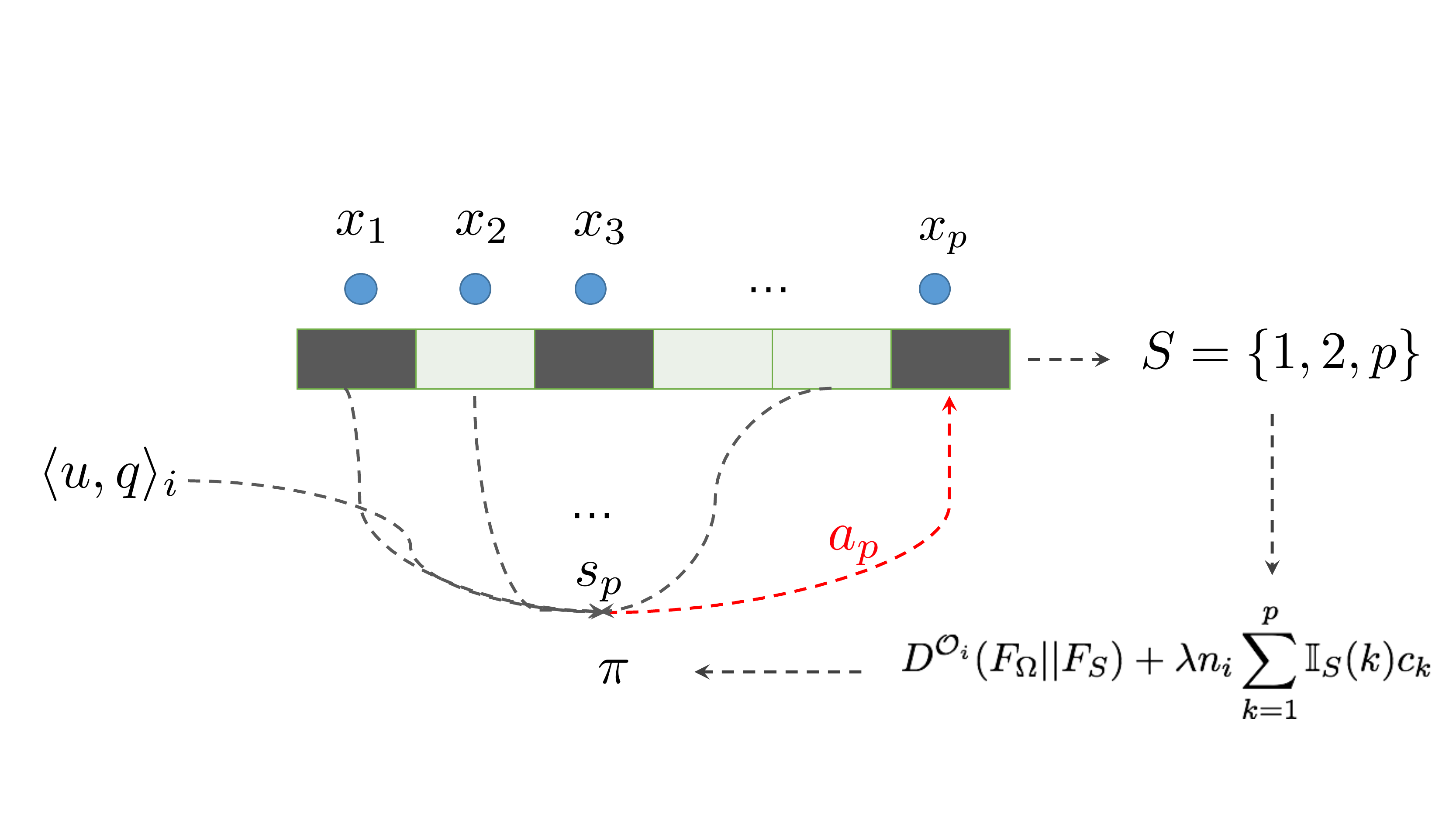}
	\caption{An example of the pruning procedure via reinforcement learning. Our goal is to select a subset of factors to optimize the objective defined in Equation \ref{eq:generalobjective}.}
	\label{fig:pruningillustration}
	\vspace{-5pt}
\end{figure}

To reduce the action space, we introduce a fixed factor sequence so that a policy can sequentially determine the corresponding utilization of factors. Formally, for each user-query $\langle u,q\rangle_i$ request, the vector function $\mathbb{I}_{\bm{\theta}}$ can be determined in $p$ steps, where in the $k$-th step ($1\le k \le p$), we need to decide whether the $k$-th factor should be applied to the ranking function or not  for this certain request, i.e.,  $\bm{a}_k \in \mathcal{A}=\{\text{Skip}, \text{Keep}\}$ is the action taken at step $k$ and the $\mathcal{A}$ is the action space. $\bm{a}_k$ is obtained through a policy
\begin{align}
\bm{a}_k = \bm{\pi}(\bm{s}_k|\bm{\theta}),
\end{align}
where $\bm{s}_k$ is the state representation of $k$-th step. Then we can get
\begin{align}
\mathbb{I}_{\bm{\theta}}(k) = \begin{cases}
0 &\text{if $\bm{a}_k=$ Skip}\\
1 &\text{if $\bm{a}_k=$ Keep}
\end{cases}
\end{align}
After $p$ steps, $\mathbb{I}_{\bm{\theta}}$ is determined and so is ranking permutation $\sigma^{\mathbb{I}_{\bm{\theta}}}$. Then we can directly calculate the loss $\mathcal{L}(\langle u,q \rangle_i, \mathcal{O}_i, f  \mid \bm{\theta})$ to evaluate the result of selected actions, which can be further used to define the total reward of the actions generated in $p$ steps during the episode. See Figure \ref{fig:pruningillustration} for illustration. The key idea lies in the state design (base on which the action is generated), the reward design (how to evaluate each action) and the optimization method for this reinforcement learning problem (how to find the optimal policy).

\subsection{The State and Reward Design}
The optimal policy should generalize over the state space, and the optimal actions for an episode only depend on the $\langle u,q \rangle_i$ request, so ideally, the state can be designed as 
\begin{align}
\bm{s}_k = (v_{\langle u,q \rangle_i}, k) \in R^{l+1}
\end{align}
where $v_{\langle u,q \rangle_i} \in R^l $ is the representations for the user-query pair $\langle u,q \rangle_i$. The corresponding reward $r_k$ is then defined as 
\begin{align}
r_k =  \begin{cases}
0 &\text{$1\le k<p$}\\
-\mathcal{L}(\langle u,q \rangle_i, \mathcal{O}_i, f  \mid \bm{\theta}) &\text{$k=p$}
\end{cases}
\end{align}
The agent is designed to obtain a reward of $0$ when the episode is not terminated, i.e., $1\le k<p$, and a reward of -$\mathcal{L}(\langle u,q \rangle_i, \mathcal{O}_i, f  \mid \bm{\theta}) $ when the episode ends, like the goal-directed tasks.  By above definition and assigning $\gamma$ to 1, then we can conclude that the objective in this reinforcement learning problem is exactly the negative of the objective in Equation \ref{Eq:linearobjective}:
\begin{align}
\label{Eq:consistent}
 \mathfrak{R}(\bm{\tau}) &=  \sum_{k=1}^{p} \gamma^{k-1} r_{k}=-\mathcal{L}(\langle u,q \rangle_i, \mathcal{O}_i, f  \mid \bm{\theta}) 
\end{align}
This means that maximizing $ \mathfrak{R}(\bm{\tau})$ can directly minimize $\mathcal{L}$, allowing us to find the optimal solution of $\mathcal{L}$ with the power of deep reinforcement learning.

However, empirically there are two issues that make learning the optimal policy for above reinforcement learning problem difficult. One is that the reward is sparse over states, known as the \emph{sparse feedback problem} ~\cite{kulkarni2016hierarchical}. The other one is that the reward itself ($\mathcal{L}$) distributes widely in the continuous space, making the critic model difficult to converge. Inspired by the \emph{reward shaping}~\cite{ng1999policy} technique, we consider to slightly change the representations of states and rewards, to alleviate the above issues.

We firstly initialize $\mathbb{I}^{'}_{\bm{\theta}}=[1,1,...,1] \in R^p$ as an all-one vector, and at the step $k$ update $\mathbb{I}^{'}_{\bm{\theta}}$ as
\begin{align}
\mathbb{I}^{'}_{\bm{\theta}}(t|k) = \begin{cases}
\mathbb{I}_{\bm{\theta}}(t) & 1\le t < k\\
1 & k \le t \le p
\end{cases}
\end{align}
Then we extend our state vector to
\begin{align}
\label{Eq:state}
\bm{s}_k = (v_{\langle u,q \rangle_i},k, \mathbb{I}^{'}_{\bm{\theta}}(\cdot|k)) \in R^{l+p+1}.
\end{align}

\begin{algorithm}[t]
	\caption{RankCFS}
	\label{Alg:acalgorithm}
	\begin{algorithmic}[1]
		\REQUIRE ~~\\
		\makebox[8em][l]{$D$:} Training data set $\mathcal{D}=\left\{\left(\langle u,q \rangle_i,\mathcal{O}_{i}\right)\right\}_{i=1}^N$ \\
		\makebox[8em][l]{$f$:} The ranking function\\
		\makebox[8em][l]{$\gamma, \lambda,\beta,r_c,T_{max}$:} Parameters of the algorithm\\
		\ENSURE ~~\\
		\makebox[8em][l]{$\bm{\theta}$:} Parameters of actor model
		
		\STATE Initialize the actor network params $\bm{\theta}$ and the critic network params $\bm{\mu}$
		\STATE $T \leftarrow 1$
		\REPEAT
		\FOR {each $\left(\langle u,q \rangle_i,\mathcal{O}_{i}\right) \in \mathcal{D}$ (For each page view)}
		\STATE $T \leftarrow T+1$
		\STATE Initialized the initial state $\bm{s}_1$ as in Eq. \ref{Eq:state}
		\FOR {$k=1,2,\dots,p$}
		\STATE Taking action $\bm{a}_k \in \{\text{Skip, Keep}\}$ on the $k$-th factor based on $\pi_{\bm{\theta}}(\bm{s}_k)$, observe $r_k$ and $\bm{s}_{k+1}$.
		\STATE Cache the tuple $(\bm{s}_k,\bm{a}_k,r_k,\bm{s}_{k+1})$
		\ENDFOR 
		\STATE $\mathfrak{R} \leftarrow 0$
		\FOR {$k=p,p-1,\dots,1$}
		\STATE $\mathfrak{R} \leftarrow r_k + \gamma \mathfrak{R}$
		\STATE $\bm{\theta} \leftarrow  \text{Adam}(\bm{\theta}, \nabla_{\bm{\theta}} \log{\pi_{\bm{\theta}}(\bm{a}_k|\bm{s}_k)}(\mathfrak{R} - V^{\pi_{\bm{\theta}}}(\bm{s}_t|\mu)))$
		\STATE $\bm{\mu} \leftarrow \text{Adam}(\bm{\mu}, (\mathfrak{R}-V^{\pi_{\bm{\theta}}}(\bm{s}_k|\bm{\mu})) \nabla_{\bm{\mu}} V^{\pi_{\bm{\theta}}}(\bm{s}_k|\bm{\mu}))$
		\ENDFOR
		\ENDFOR
		\UNTIL{$T>T_{max}$}
	\end{algorithmic}
\end{algorithm}

Thus our state memorizes the decisions made before during an episode. At each step $k$,  the reward is calculated based on $ \mathbb{I}^{'}_{\bm{\theta}}(\cdot|k)$, i.e., at each step it is pre-evaluated for the decisions made so far, assuming the rest decisions are all ones by default. For each reward $r_k$, we decompose it into the effectiveness part $\mathcal{T}(\bm{s}_k,\bm{a}_k)$ and the efficiency part $\mathcal{G}(\bm{s}_k,\bm{a}_k)$, i.e., $r_k=\mathcal{T}(\bm{s}_k,\bm{a}_k)+\mathcal{G}(\bm{s}_k,\bm{a}_k)$. For the efficiency part, we simply add a penalty when keeping the $k$-th factor as 
\begin{align}
{\mathcal{G}}(\bm{s}_k,\bm{a}_k) = \begin{cases}
0 &\text{if $\bm{a}_k=$ Skip}\\
-\lambda n_i c_k &\text{if $\bm{a}_k=$ Keep}
\end{cases}
\end{align}
This part is consistent with the Equation \ref{Eq:linearobjective}. For the effectiveness part, we choose to give a constant penalty if the ranking loss under $\mathbb{I}_{\bm{\theta}}^{'}$ exceeds a pre-defined threshold as
 \begin{align}
\mathcal{T}(\bm{s}_k,\bm{a}_k) = \begin{cases}
-r_{c} & D^{\mathcal{O}_i}(\sigma_{\Omega}||\sigma_i^{\mathbb{I}_{\bm{\theta}}^{'}}) > \beta\\
0 &\text{otherwise.}
\end{cases}
\end{align}
rather than $-D^{\mathcal{O}_i}(\sigma_{\Omega}||\sigma_i^{\mathbb{I}_{\bm{\theta}}^{'}})$ itself shown in Equation \ref{Eq:linearobjective}. By such design we could help the critic distinguish \emph{bad} and \emph{good} ranking result much easier. Moreover, we could avoid generating poor ranking performance with a large penalty $r_c$.

\subsection{Learning the Policy}

After transforming the original problem into a reinforcement learning one, we could then apply any reinforcement learning methods. In this paper, we choose the well-known policy gradient method with actor-critic models as described in~\cite{mnih2016asynchronous} and we call it RankCFS. It is worthing noting that, the difficulty of the original optimization problem does not decrease with the introduction of reinforcement learning techniques. The RL-based approach here acts as a solver whose solution space contains the optimal, and provides an efficient searching path to the optimal through trial-and-error methods. 

Algorithm \ref{Alg:acalgorithm} shows the training details. The data of page views in the on-line search system $\mathcal{D}=\left\{\left(\langle u,q \rangle_i,\mathcal{O}_{i}\right)\right\}_{i=1}^N$, the reward discount factor $\gamma$, the parameters used in the reward definition $\lambda,\beta,r_c$ and the maximal number of training step $T_{max}$ are given as the input of the algorithm. The parameter of the actor network $\bm{\theta}$ is the output of the algorithm. We firstly initialize the parameters of the actor and critic network, as well as the step counter $T$, as in Line 1 and 2. The training phase starts with the iteration of the each page view, with which an episode will be generated during the Line 6-10. Then standard policy gradient is conducted in Line 14-15, where the tuple $(\bm{s}_k,\bm{a}_k,r_k,\bm{s}_{k+1})$ is organized in the backward way so the discounted cumulative reward $\mathfrak{R}$ can be updated incrementally as in Line 13. The training process ends when the number of steps exceeds the given threshold $T_{max}$.


\section{Experimental Results}
In this section, we provide empirical results of our approaches in off-line evaluation and commercial on-line evaluation.  We show the results of off-line settings in order to provide a way to justify our algorithm. Then, we test our method in a real on-line commercial web search engine to reveal the performance improvement with respect to the resource consumption. At last, we demonstrate the performance of our method in Singes' Day shopping festival.

\subsection{Off-line Comparison}
\begin{figure}[t]
	\centering
	\includegraphics[width=0.8\linewidth, trim=25 25 25 25,clip ]{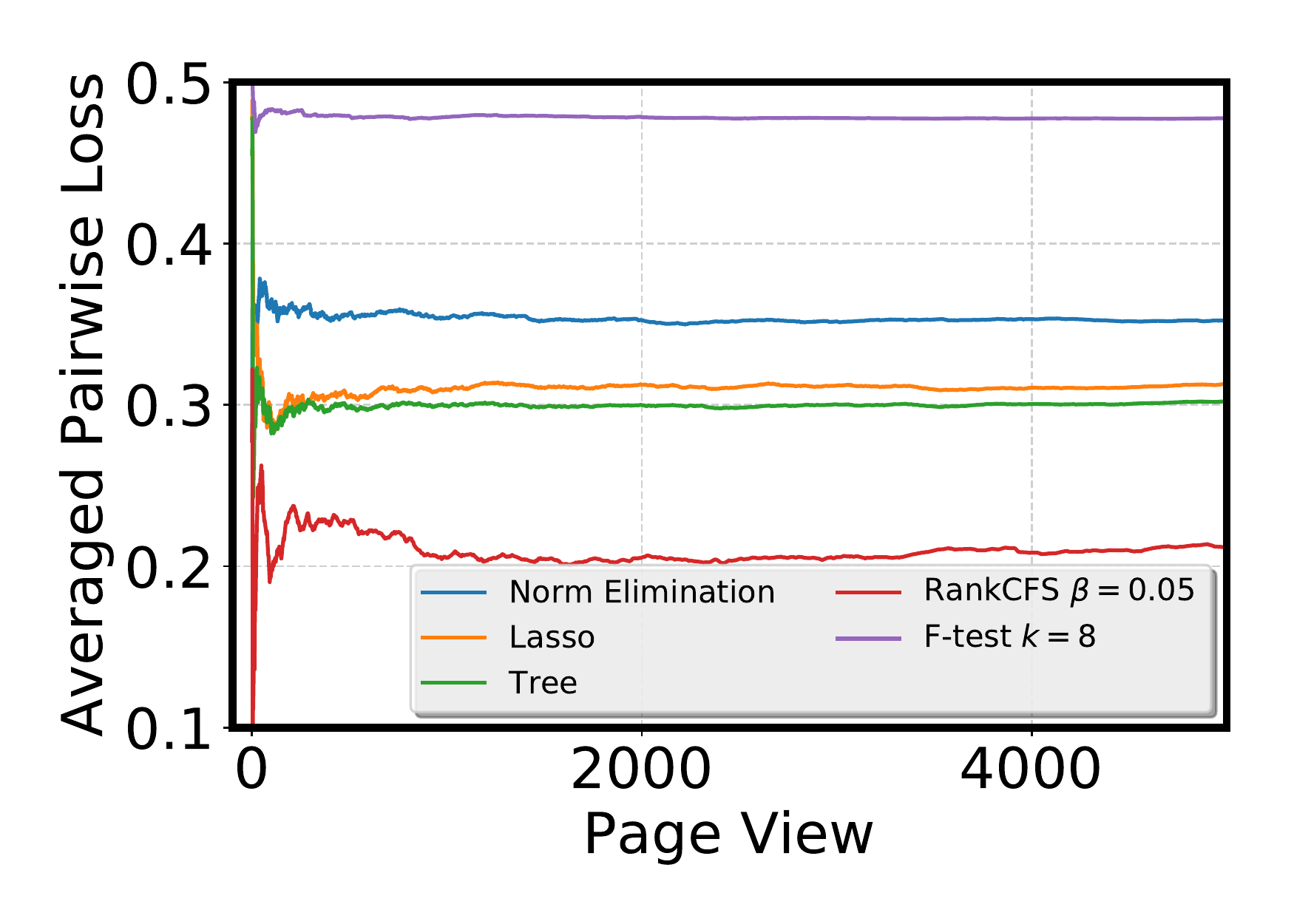}
	\caption{Pairwise loss vs Page view in off-line comparison. Lower is better.}
	\label{fig:offlineresultsloss}
\end{figure}
\begin{figure}[t]
	\centering
	\includegraphics[width=0.8\linewidth,  trim=25 25 25 25,clip]{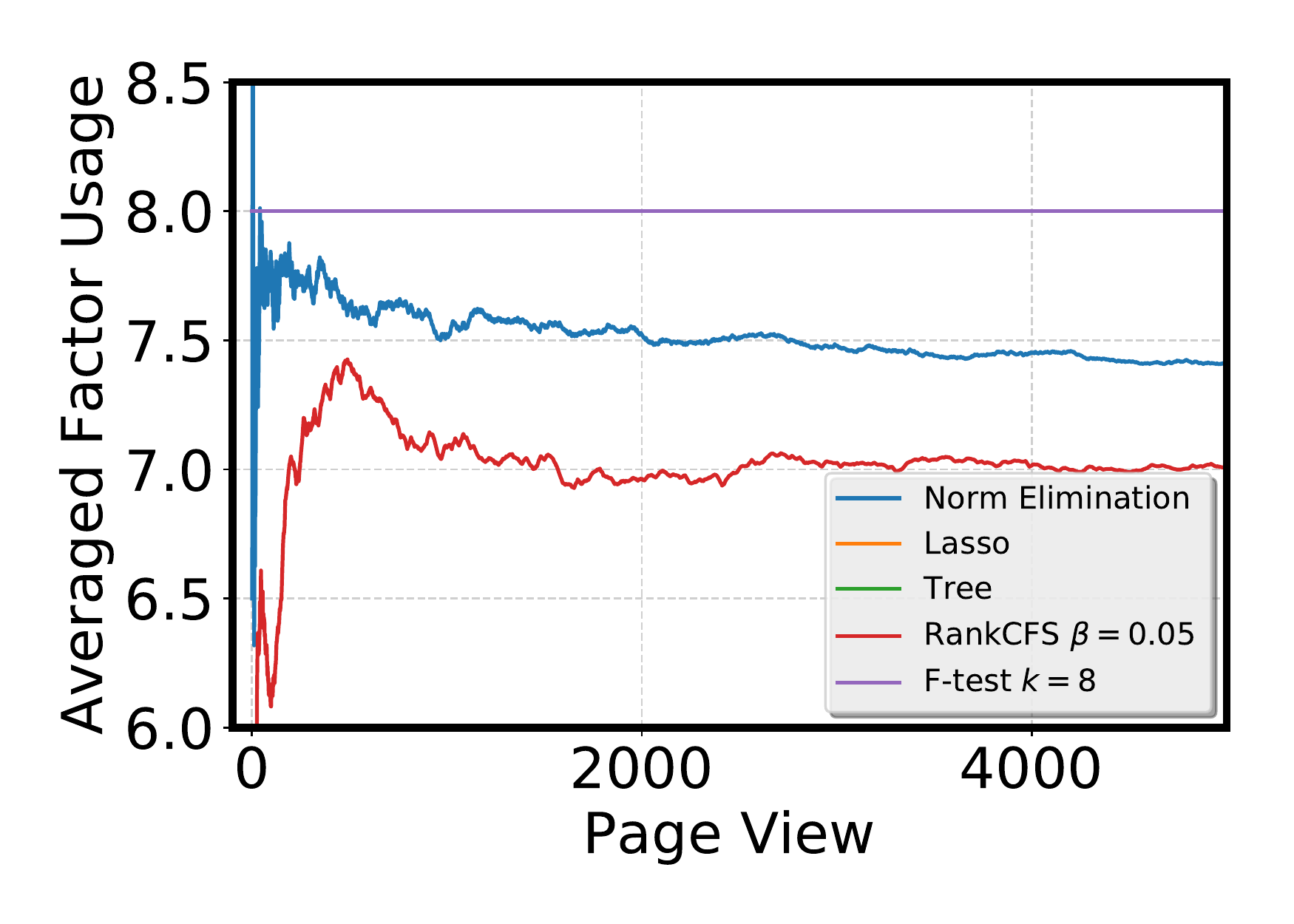}
	\caption{Factor usage vs Page view in off-line comparison. Lasso, Tree-based method and F-test k=8 have the same averaged factor usage at $8$.  Lower is better.}
	\label{fig:offlineresultsfactorusage}
\end{figure}

\begin{figure}[t]
	\centering
	\includegraphics[width=0.8\linewidth,  trim=25 25 25 25,clip]{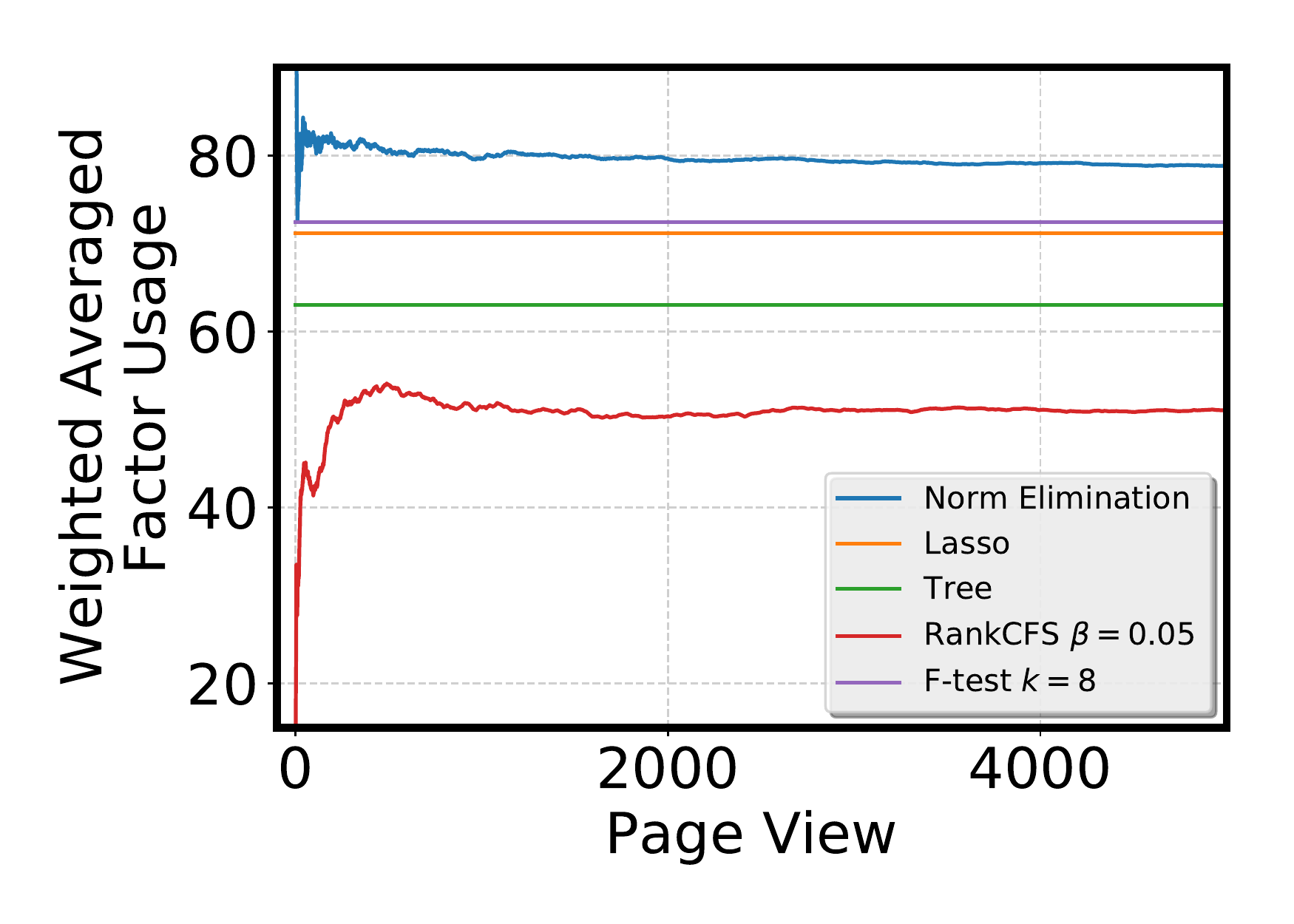}
	\caption{Weighted averaged factor usage vs Page view in off-line comparison. Lower is better. }
	\label{fig:offlineresultsfactorusage}
\end{figure}

In this subsection, we compare our method with norm elimination method, $l_1$-based feature selection, tree-based feature selection and $F$-test feature selection in an off-line evaluation setting.

\textbf{Norm Elimination} is that we remove those factors whose absolute values of weights are less than a positive constant $\epsilon$. 

\textbf{$l_1$-based feature selection} is a model-based feature selection method such that it selects factor according to the $l_1$ regularizer. The basic idea is that to eliminate those factors whose corresponding $l_1$ coefficients are zero. Since this method must be based on a supervised machine learning model, we need to convert our ranking problem into a supervised one. We define the training data set as following:
let the label $l^{i,j}=f(x^{i,j}_1,x^{i,j}_2,\dots,x^{i,j}_p)=\sum_{k=1}^p w_k^{i} x^{i,j}_k$ and corresponding factor vector $\bm{x}^{i,j}$, thus the set of training examples has the form $\mathcal{D_T}=\{l^{i,j}, \bm{x}^{i,j}\}$ such that $j=1,\dots, n_i$ and $i=1,\dots, N$. Therefore, we can train a regressor via training set $\mathcal{D_T}$ and select the factors base on the trained model. We adopt the Lasso as our comparison method.

\begin{table}[t]
	\caption{Results Summary}
	\begin{tabular}{c|ccc}
		\hline 
		\multirow{2}{*}{Algorithm} & Averaged & Averaged& Weighted  \\ 
		& Pairwise Loss& factor Usage& factor Usage\\
		\hline 
		F-test $k=8$ & $0.47$ & $8$ &$72.44$ \\ 
		F-test $k=11$ & $0.40$ & $11$ & $84.63$\\ 
		F-test $k=14$ & $0.29$ & $14$ & $107.93$\\ 
		Norm Elimination & $0.3$5 & $7.41$ & $78.84$\\ 
		Lasso & $0.31$ & $8$ &$71.16$\\ 
		Tree-based & $0.3$ & $7$& $63$\\ 
		RankCFS $\beta=0.05$ & $0.21$ & $7.01$& $51.06$\\ 
		RankCFS $\beta=0.15$ & $0.27$ & $9.07$ &$67.40$\\ 
		RankCFS $\beta=0.25$ & $0.25$ & $8.4$ &$63.7$2 \\ 
		\hline 
	\end{tabular} 
	\label{table:offlienresults}
\end{table}

\textbf{Tree-based feature selection} is similar to the $l_1$-based feature selection and the difference is  that we replace the Lasso model with a non-linear regression tree model.

\textbf{$F$-test feature selection} is a model-free feature selection method that selects top $k$ factors based on $F$-test scores.

\textbf{Rank Contextual Factor selection (RankCFS)}  algorithm is out actor-critic method which is capable of adjusting the usage of factor by contexts. 

For $l_1$-based feature selection, Tree-based feature selection and $F$-test feature selection, we adopt their implementations in  scikit-learn~\cite{scikit-learn}. We implement RankCFS with Tensorflow~\cite{abadi2016tensorflow}.
For the optimal ranking model in the off-line evaluation, we select one of linear ranking models of Taobao.com and treat it as a black box so that the input and output of the ranking model are merely considered during the experimental process. We set the constant $\epsilon=0.1$ in the Norm Elimination method; the constant that multiplies the $l_1$ term $\alpha$ equals $0.05$ in the Lasso model; We choose the default ExtraTreeRegressor in scikit-learn package as our tree model; The actor and critic are construct by two deep neural networks (DNN) with three fully connected layers, respectively. The DNN structures of actor are $266 \times 128 \times 128 \times 20$ and ones of critic are $266 \times 128\times 128 \times 1$. We adopt \textit{relu} as the activation functions for the hidden layers, Adam as our optimizer and the learning rates of actor and critic are $0.0001$ and $0.001$, receptively. 

We sample a data set with $100,000$ examples as mentioned above, then train the  $l_1$-based,  Tree-based and $F$-test approaches\footnote{It is not necessary to train Norm Elimination method since it only removes those factors whose absolute values of weights are less than a positive constant $\epsilon$.} on $50,000$ examples and test then on the rest of $50,000$ \footnote{$10,000$ page views and $10$ items in each page view.}.  For the $l_1$, Tree-based and F-test methods, the feature selection are determined after the training, that is, we use a fixed feature selection policy during the testing stage. Since our method requires to consider the computational costs of factors, the computational cost vector $\bm{c}$ is obtained from the on-line operational environment of Taobao.com. 

We test our methods on $5,000$ page views and each page view contains $10$ items so that there are $50,000$ testing examples. Then, we evaluate the averaged pairwise loss defined in Equation \ref{eq:avaragepairwiseloss} and factor usage over page views. Figure \ref{fig:offlineresultsloss}-\ref{fig:offlineresultsfactorusage} show our experimental results. Generally, the Norm Elimination method removes those factors whose absolute values are small under different contexts, therefore indicators such as loss, factor usage may vary over page views. Figure \ref{fig:offlineresultsloss} demonstrates that our RankCFS with the threshold $\beta=0.05$ outperforms all other methods in terms of pairwise loss. And in Figure \ref{fig:offlineresultsfactorusage}, it shows that the averaged factor usage of RankCFS algorithm is also close to the lowest Tree-based method. Figure \ref{fig:offlineresultsfactorusage} demonstrates the weighted factor usage, in which the weights are the corresponding computational costs. This metric is more accurate to describe the factor usage in terms of efficiency due to the variety in computational costs among factors. For example , it only considers absolute values of weights in the Norm Elimination method, while RankCFS tends to eliminate those factors with high computational costs. Although, RankCFS $\beta=0.05$ and Tree-based method have similar averaged factor usage, RankCFS $\beta=0.05$ has much lower weighted  factor usage. It is the evidence that our approach relieves the computational burden in an intelligent way and save more computational resources.
Empirically, our method is capable of exploring better solution in a combinatorial solution space with less factor usage. The F-test method suffers high pairwise loss since it is a model-free method, it is not required to consider the ranking model we adopt.
Overall, the experiments shows that our RankCFS algorithm successfully explores the function space and find an excellent approximation to the optimal ranking function. We summarized the complete experimental results in Table \ref{table:offlienresults} with varieties in parameters.  Note that RankCFS $\beta=0.25$ outperforms RankCFS $\beta=0.15$, it is possible that RankCFS $\beta=0.15$ falls into a worse local optimal solution and fails to escape from it.

\begin{figure}[t!]
	\centering
	\begin{subfigure}{1\columnwidth}
		\includegraphics[width=1\linewidth]{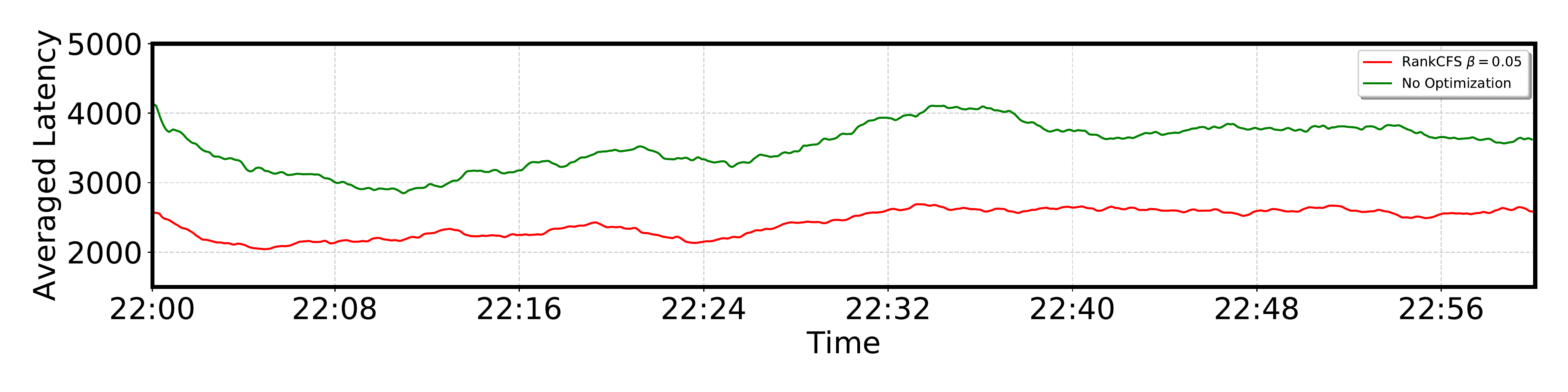}
		\caption{Latency fo regular operational environment}
		\label{fig:latencydoubleregular}
	\end{subfigure}
	\begin{subfigure}{1\columnwidth}
		\centering
		\includegraphics[width=1\linewidth]{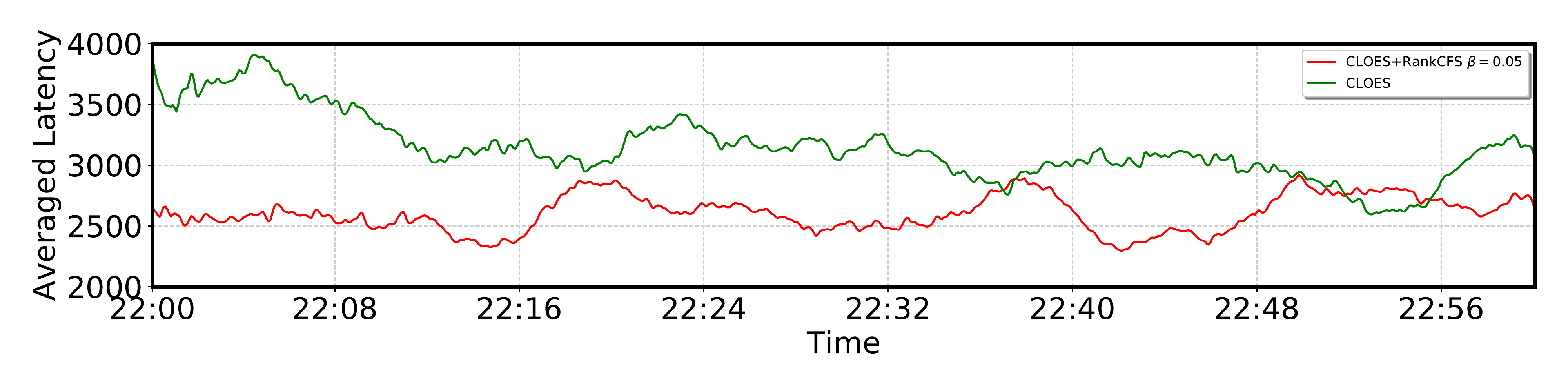}
		\caption{Latency for Singles' Day}
		\label{fig:latencydouble11}
	\end{subfigure}
	\caption{Latency in a real-world large-scale e-commerce search engine. Lower is better.}
\end{figure}

\subsection{On-line Evaluation in Operational Environment}

In this subsection and the following subsection, we present the experimental results of on-line evaluation in the real-world large-scale operational environment of Taobao.com with a standard A/B testing setting. For the on-line evaluation, we adopt the same learning structure with the off-line one, but with a more complex nonlinear optimal ranking model. The training is conducted with more than $1 \times10^9$ training samples on a distributed streaming system in an on-line learning fashion.  The system information of computers in the clusters on which we conducted our experiments is list in Table \ref{table:sysinfo}. 
\begin{table}
	\caption{System information}
\begin{tabular}{l|l}
	\hline 
	Hardware & Configuration \\ 
	\hline 
	\multirow{2}{*}{CPU} & 2x 16-core Intel(R) Xeon(R)  \\ 
	&CPU E5-2682 v4 @ 2.5GHz \\
	RAM & 256 GB \\ 
	Hyperthreading & Yes \\ 
	Networking & 10 Gbps \\
	OS & ALIOS7 Linux 3.10.0 x86\_64 \\ 
	\hline 
\end{tabular} 
	\label{table:sysinfo}
\end{table}

The search engine of Taobao.com is a complex system, processing billions of items and hundreds of millions of user queries every day. As a core system in Taobao.com, the search engine needs to respond the user queries in a timely manner. The search traffic might increase significantly during some promotional campaign such as the Singes' Day shopping festival. Therefore, the system efficiency is always an important issue. Furthermore, the system is still required to provide high quality search service to the users, leading to computational burden for the whole system.

We conduct a standard A/B test experiment in our operational environment, where roughly 6\% of the random users are select for the testing. The parameter $\beta \in \{0.25,0.15,0.05\}$ and $\lambda \in \{0.9,0.8,0.7\}$ are tuned through the GMV and search latency. The goal is to minimize the impact on the GMV as much as possible, while reducing the latency as much as we can, comparing the control group. Figure \ref{fig:latencydoubleregular} shows the best result with $\beta=0.05$ and $\lambda=0.9$. Our method saves approximately $40\%$ average search latency, comparing to the control group. For the max search latency, out algorithm reduce roughly $25\%$ latency. The system performance (GMV) is almost the same or little lower ($0\%$ to $0.5\%$) as the control group.

\subsection{Singles' Day Evaluation}

Alibaba Singles' Day shopping festival is one of the biggest shopping extravaganza around the world and is the Chinese version of Black Friday. In 2017, by the end of day (November 11), sales hit a new record of \$25.3 billion, more than $40\%$ higher than sales on Singles' Day 2016 and it attracts over hundreds of millions users from more than 200 different countries. The infrastructure system manages to handle 0.325 millions orders per second at peak\footnote{\url{https://techcrunch.com/2017/11/11/alibaba-smashes-its-singles-day-record/}}. The e-commerce search system played a crucial role in this event.

In November 11th, the search traffic burden of the e-commerce search engine abruptly increases by multiple times as much as in a regular day. On the one hand, the e-commerce search engine faces the high traffic challenge, which might lead to system degradation. On the other hand, it is still crucial to provide high search accuracy even during shopping festival. 

In the event, our method collaborates previous work \cite{Liu:2017:CRO:3097983.3098011} in order to maximize optimization at the search engine system level. The algorithm called CLOES mainly focuses on optimizing the number of items in the ranking process via cascading model, while our method concentrates on optimizing the set of ranking factors during the ranking process. Thus, both of approaches are able to apply on the search engine simultaneously. We use CLOSE approach as our control group, CLOES+RankCFS as the experimental group, in which the parameter $\beta=0.05$\footnote{Due to the limited on-line resources, we are only able to use CLOES as our control group.}.  Figure \ref{fig:latencydouble11} depicts the average latency change during the experiments, in which our method averagely saves $20\%$ more latency on the basis of CLOES. And also, our method saves approximately $33\%$ peak latency on the basis of CLOES. The system performance (GMV) is almost the same as the CLOSE method.

Our method and CLOES collaborate together in the very day of Singes' Day 2017, and succeed in providing a much better search performance than previous year. 

\section{Conclusion and Future Work}
\label{sec:conclusion}
In this paper, we thoroughly investigate the effectiveness and efficiency issues in a real-world large-scale e-commerce search system and propose an intelligent optimization solution by reinforcement learning method. We formally defined the learning to rank problem in an e-commerce scenario and characterize the effectiveness and efficiency, which is a NP-hard problem. Then, we convert the problem into a reinforcement learning problem by the reward design and solve it by the actor-critic method. We empirically test our method in off-line and on-line evaluation scenarios,  demonstrating our method is a practical solution in a real-world large-scale e-commerce search system. In future, we plan on finding other ways to optimize the system engine such as memory usage, load balancing, combined with search latency. Moreover, the DNN network representation in our current setting is not an end-to-end solution so that the end-to-end representation solution (i.e. pointer network \cite{vinyals2015pointer}) will be considered.

\bibliographystyle{ACM-Reference-Format}
\bibliography{rl_perfopt}


\begin{thebibliography}{35}


\ifx \showCODEN    \undefined \def \showCODEN     #1{\unskip}     \fi
\ifx \showDOI      \undefined \def \showDOI       #1{#1}\fi
\ifx \showISBNx    \undefined \def \showISBNx     #1{\unskip}     \fi
\ifx \showISBNxiii \undefined \def \showISBNxiii  #1{\unskip}     \fi
\ifx \showISSN     \undefined \def \showISSN      #1{\unskip}     \fi
\ifx \showLCCN     \undefined \def \showLCCN      #1{\unskip}     \fi
\ifx \shownote     \undefined \def \shownote      #1{#1}          \fi
\ifx \showarticletitle \undefined \def \showarticletitle #1{#1}   \fi
\ifx \showURL      \undefined \def \showURL       {\relax}        \fi
\providecommand\bibfield[2]{#2}
\providecommand\bibinfo[2]{#2}
\providecommand\natexlab[1]{#1}
\providecommand\showeprint[2][]{arXiv:#2}

\bibitem[\protect\citeauthoryear{Abadi, Barham, Chen, Chen, Davis, Dean, Devin,
  Ghemawat, Irving, Isard, et~al\mbox{.}}{Abadi et~al\mbox{.}}{2016}]%
        {abadi2016tensorflow}
\bibfield{author}{\bibinfo{person}{Mart{\'\i}n Abadi}, \bibinfo{person}{Paul
  Barham}, \bibinfo{person}{Jianmin Chen}, \bibinfo{person}{Zhifeng Chen},
  \bibinfo{person}{Andy Davis}, \bibinfo{person}{Jeffrey Dean},
  \bibinfo{person}{Matthieu Devin}, \bibinfo{person}{Sanjay Ghemawat},
  \bibinfo{person}{Geoffrey Irving}, \bibinfo{person}{Michael Isard},
  {et~al\mbox{.}}} \bibinfo{year}{2016}\natexlab{}.
\newblock \showarticletitle{TensorFlow: A System for Large-Scale Machine
  Learning.}. In \bibinfo{booktitle}{\emph{OSDI}}, Vol.~\bibinfo{volume}{16}.
  \bibinfo{pages}{265--283}.
\newblock


\bibitem[\protect\citeauthoryear{Bello, Pham, Le, Norouzi, and Bengio}{Bello
  et~al\mbox{.}}{2016}]%
        {bello2016neural}
\bibfield{author}{\bibinfo{person}{Irwan Bello}, \bibinfo{person}{Hieu Pham},
  \bibinfo{person}{Quoc~V Le}, \bibinfo{person}{Mohammad Norouzi}, {and}
  \bibinfo{person}{Samy Bengio}.} \bibinfo{year}{2016}\natexlab{}.
\newblock \showarticletitle{Neural combinatorial optimization with
  reinforcement learning}.
\newblock \bibinfo{journal}{\emph{arXiv preprint arXiv:1611.09940}}
  (\bibinfo{year}{2016}).
\newblock


\bibitem[\protect\citeauthoryear{Benbouzid, Busa-Fekete, and
  K{\'e}gl}{Benbouzid et~al\mbox{.}}{2012}]%
        {busa2012fast}
\bibfield{author}{\bibinfo{person}{Djalel Benbouzid},
  \bibinfo{person}{R{\"o}bert Busa-Fekete}, {and} \bibinfo{person}{Bal{\'a}zs
  K{\'e}gl}.} \bibinfo{year}{2012}\natexlab{}.
\newblock \showarticletitle{Fast classification using sparse decision DAGs}. In
  \bibinfo{booktitle}{\emph{Proceedings of the 29th International Conference on
  International Conference on Machine Learning}}. Omnipress,
  \bibinfo{pages}{747--754}.
\newblock


\bibitem[\protect\citeauthoryear{Bhatnagar, Sutton, Ghavamzadeh, and
  Lee}{Bhatnagar et~al\mbox{.}}{2009}]%
        {bhatnagar2009natural}
\bibfield{author}{\bibinfo{person}{Shalabh Bhatnagar},
  \bibinfo{person}{Richard~S Sutton}, \bibinfo{person}{Mohammad Ghavamzadeh},
  {and} \bibinfo{person}{Mark Lee}.} \bibinfo{year}{2009}\natexlab{}.
\newblock \showarticletitle{Natural actor--critic algorithms}.
\newblock \bibinfo{journal}{\emph{Automatica}} \bibinfo{volume}{45},
  \bibinfo{number}{11} (\bibinfo{year}{2009}), \bibinfo{pages}{2471--2482}.
\newblock


\bibitem[\protect\citeauthoryear{Bourdev and Brandt}{Bourdev and
  Brandt}{2005}]%
        {bourdev2005robust}
\bibfield{author}{\bibinfo{person}{Lubomir Bourdev} {and}
  \bibinfo{person}{Jonathan Brandt}.} \bibinfo{year}{2005}\natexlab{}.
\newblock \showarticletitle{Robust object detection via soft cascade}. In
  \bibinfo{booktitle}{\emph{Computer Vision and Pattern Recognition, 2005. IEEE
  Computer Society Conference on}}, Vol.~\bibinfo{volume}{2}. IEEE,
  \bibinfo{pages}{236--243}.
\newblock


\bibitem[\protect\citeauthoryear{Burges, Shaked, Renshaw, Lazier, Deeds,
  Hamilton, and Hullender}{Burges et~al\mbox{.}}{2005}]%
        {Burges2005Learning}
\bibfield{author}{\bibinfo{person}{Chris Burges}, \bibinfo{person}{Tal Shaked},
  \bibinfo{person}{Erin Renshaw}, \bibinfo{person}{Ari Lazier},
  \bibinfo{person}{Matt Deeds}, \bibinfo{person}{Nicole Hamilton}, {and}
  \bibinfo{person}{Greg Hullender}.} \bibinfo{year}{2005}\natexlab{}.
\newblock \showarticletitle{Learning to Rank Using Gradient Descent}. In
  \bibinfo{booktitle}{\emph{Proceedings of the 22nd International Conference on
  Machine Learning}}. \bibinfo{publisher}{ACM}, \bibinfo{address}{New York, NY,
  USA}, \bibinfo{pages}{89--96}.
\newblock
\showISBNx{1-59593-180-5}


\bibitem[\protect\citeauthoryear{Burges}{Burges}{2010}]%
        {Burges2010From}
\bibfield{author}{\bibinfo{person}{Chris~J.C. Burges}.}
  \bibinfo{year}{2010}\natexlab{}.
\newblock \bibinfo{booktitle}{\emph{From RankNet to LambdaRank to LambdaMART:
  An Overview}}.
\newblock \bibinfo{type}{{T}echnical {R}eport}.
\newblock
\urldef\tempurl%
\url{https://www.microsoft.com/en-us/research/publication/from-ranknet-to-lambdarank-to-lambdamart-an-overview/}
\showURL{%
\tempurl}


\bibitem[\protect\citeauthoryear{Cao, Qin, Liu, Tsai, and Li}{Cao
  et~al\mbox{.}}{2007}]%
        {learning-to-rank-from-pairwise-approach-to-listwise-approach}
\bibfield{author}{\bibinfo{person}{Zhe Cao}, \bibinfo{person}{Tao Qin},
  \bibinfo{person}{Tie-Yan Liu}, \bibinfo{person}{Ming-Feng Tsai}, {and}
  \bibinfo{person}{Hang Li}.} \bibinfo{year}{2007}\natexlab{}.
\newblock \bibinfo{booktitle}{\emph{Learning to Rank: From Pairwise Approach to
  Listwise Approach}}.
\newblock \bibinfo{type}{{T}echnical {R}eport}.
\newblock
\urldef\tempurl%
\url{https://www.microsoft.com/en-us/research/publication/learning-to-rank-from-pairwise-approach-to-listwise-approach/}
\showURL{%
\tempurl}


\bibitem[\protect\citeauthoryear{Cooper, Gey, and Dabney}{Cooper
  et~al\mbox{.}}{1992}]%
        {Cooper1992Probabilistic}
\bibfield{author}{\bibinfo{person}{William~S. Cooper},
  \bibinfo{person}{Fredric~C. Gey}, {and} \bibinfo{person}{Daniel~P. Dabney}.}
  \bibinfo{year}{1992}\natexlab{}.
\newblock \showarticletitle{Probabilistic Retrieval Based on Staged Logistic
  Regression}. In \bibinfo{booktitle}{\emph{Proceedings of the 15th Annual
  International ACM SIGIR Conference on Research and Development in Information
  Retrieval}}. \bibinfo{publisher}{ACM}, \bibinfo{address}{New York, NY, USA},
  \bibinfo{pages}{198--210}.
\newblock
\showISBNx{0-89791-523-2}


\bibitem[\protect\citeauthoryear{Davis, Mallat, and Avellaneda}{Davis
  et~al\mbox{.}}{1997}]%
        {davis1997adaptive}
\bibfield{author}{\bibinfo{person}{Geoff Davis}, \bibinfo{person}{Stephane
  Mallat}, {and} \bibinfo{person}{Marco Avellaneda}.}
  \bibinfo{year}{1997}\natexlab{}.
\newblock \showarticletitle{Adaptive greedy approximations}.
\newblock \bibinfo{journal}{\emph{Constructive Approximation}}
  \bibinfo{volume}{13}, \bibinfo{number}{1} (\bibinfo{year}{1997}),
  \bibinfo{pages}{57--98}.
\newblock


\bibitem[\protect\citeauthoryear{Freund, Iyer, Schapire, and Singer}{Freund
  et~al\mbox{.}}{2003}]%
        {Freund2003An}
\bibfield{author}{\bibinfo{person}{Yoav Freund}, \bibinfo{person}{Raj Iyer},
  \bibinfo{person}{Robert~E Schapire}, {and} \bibinfo{person}{Yoram Singer}.}
  \bibinfo{year}{2003}\natexlab{}.
\newblock \showarticletitle{An efficient boosting algorithm for combining
  preferences}.
\newblock \bibinfo{journal}{\emph{Journal of Machine Learning Research}}
  \bibinfo{volume}{4}, \bibinfo{number}{6} (\bibinfo{year}{2003}),
  \bibinfo{pages}{170--178}.
\newblock


\bibitem[\protect\citeauthoryear{Geng, Liu, Qin, and Li}{Geng
  et~al\mbox{.}}{2007}]%
        {Geng:2007:FSR:1277741.1277811}
\bibfield{author}{\bibinfo{person}{Xiubo Geng}, \bibinfo{person}{Tie-Yan Liu},
  \bibinfo{person}{Tao Qin}, {and} \bibinfo{person}{Hang Li}.}
  \bibinfo{year}{2007}\natexlab{}.
\newblock \showarticletitle{Feature Selection for Ranking}. In
  \bibinfo{booktitle}{\emph{Proceedings of the 30th Annual International ACM
  SIGIR Conference on Research and Development in Information Retrieval}}.
  \bibinfo{publisher}{ACM}, \bibinfo{address}{New York, NY, USA},
  \bibinfo{pages}{407--414}.
\newblock
\showISBNx{978-1-59593-597-7}


\bibitem[\protect\citeauthoryear{Guyon and Elisseeff}{Guyon and
  Elisseeff}{2003}]%
        {guyon2003introduction}
\bibfield{author}{\bibinfo{person}{Isabelle Guyon} {and}
  \bibinfo{person}{Andr{\'e} Elisseeff}.} \bibinfo{year}{2003}\natexlab{}.
\newblock \showarticletitle{An introduction to variable and feature selection}.
\newblock \bibinfo{journal}{\emph{Journal of machine learning research}}
  \bibinfo{volume}{3}, \bibinfo{number}{Mar} (\bibinfo{year}{2003}),
  \bibinfo{pages}{1157--1182}.
\newblock


\bibitem[\protect\citeauthoryear{Kira and Rendell}{Kira and Rendell}{1992}]%
        {kira1992feature}
\bibfield{author}{\bibinfo{person}{Kenji Kira} {and} \bibinfo{person}{Larry~A
  Rendell}.} \bibinfo{year}{1992}\natexlab{}.
\newblock \showarticletitle{The feature selection problem: traditional methods
  and a new algorithm}. In \bibinfo{booktitle}{\emph{Proceedings of the 10th
  National Conference on Artificial Intelligence}}. AAAI Press,
  \bibinfo{pages}{129--134}.
\newblock


\bibitem[\protect\citeauthoryear{Kober and Peters}{Kober and Peters}{2011}]%
        {kober2009policy}
\bibfield{author}{\bibinfo{person}{Jens Kober} {and} \bibinfo{person}{Jan
  Peters}.} \bibinfo{year}{2011}\natexlab{}.
\newblock \showarticletitle{Policy search for motor primitives in robotics}.
\newblock \bibinfo{journal}{\emph{Machine Learning}} \bibinfo{volume}{1},
  \bibinfo{number}{84} (\bibinfo{year}{2011}), \bibinfo{pages}{171--203}.
\newblock


\bibitem[\protect\citeauthoryear{Kulkarni, Narasimhan, Saeedi, and
  Tenenbaum}{Kulkarni et~al\mbox{.}}{2016}]%
        {kulkarni2016hierarchical}
\bibfield{author}{\bibinfo{person}{Tejas~D Kulkarni}, \bibinfo{person}{Karthik
  Narasimhan}, \bibinfo{person}{Ardavan Saeedi}, {and} \bibinfo{person}{Josh
  Tenenbaum}.} \bibinfo{year}{2016}\natexlab{}.
\newblock \showarticletitle{Hierarchical Deep Reinforcement Learning:
  Integrating Temporal Abstraction and Intrinsic Motivation}.
\newblock In \bibinfo{booktitle}{\emph{Advances in Neural Information
  Processing Systems 29}}, \bibfield{editor}{\bibinfo{person}{D.~D. Lee},
  \bibinfo{person}{M.~Sugiyama}, \bibinfo{person}{U.~V. Luxburg},
  \bibinfo{person}{I.~Guyon}, {and} \bibinfo{person}{R.~Garnett}} (Eds.).
  \bibinfo{publisher}{Curran Associates, Inc.}, \bibinfo{pages}{3675--3683}.
\newblock


\bibitem[\protect\citeauthoryear{Kullback and Leibler}{Kullback and
  Leibler}{1951}]%
        {kullback1951information}
\bibfield{author}{\bibinfo{person}{Solomon Kullback} {and}
  \bibinfo{person}{Richard~A Leibler}.} \bibinfo{year}{1951}\natexlab{}.
\newblock \showarticletitle{On information and sufficiency}.
\newblock \bibinfo{journal}{\emph{The annals of mathematical statistics}}
  \bibinfo{volume}{22}, \bibinfo{number}{1} (\bibinfo{year}{1951}),
  \bibinfo{pages}{79--86}.
\newblock


\bibitem[\protect\citeauthoryear{Li, Burges, and Wu}{Li et~al\mbox{.}}{2008}]%
        {learning-to-rank-using-classification-and-gradient-boosting}
\bibfield{author}{\bibinfo{person}{Ping Li}, \bibinfo{person}{Chris~J.C.
  Burges}, {and} \bibinfo{person}{Qiang Wu}.} \bibinfo{year}{2008}\natexlab{}.
\newblock \showarticletitle{Learning to Rank Using Classification and Gradient
  Boosting}, In \bibinfo{booktitle}{Advances in Neural Information Processing
  Systems 20}.
\newblock


\bibitem[\protect\citeauthoryear{Liu, Yu, Jiang, and Zhou}{Liu
  et~al\mbox{.}}{2008}]%
        {liu2008tefe}
\bibfield{author}{\bibinfo{person}{Li-Ping Liu}, \bibinfo{person}{Yang Yu},
  \bibinfo{person}{Yuan Jiang}, {and} \bibinfo{person}{Zhi-Hua Zhou}.}
  \bibinfo{year}{2008}\natexlab{}.
\newblock \showarticletitle{TEFE: A Time-Efficient Approach to Feature
  Extraction}. In \bibinfo{booktitle}{\emph{Proceedings of the 8th IEEE
  International Conference on Data Mining}}. IEEE Computer Society,
  \bibinfo{pages}{423--432}.
\newblock


\bibitem[\protect\citeauthoryear{Liu, Xiao, Ou, and Si}{Liu
  et~al\mbox{.}}{2017}]%
        {Liu:2017:CRO:3097983.3098011}
\bibfield{author}{\bibinfo{person}{Shichen Liu}, \bibinfo{person}{Fei Xiao},
  \bibinfo{person}{Wenwu Ou}, {and} \bibinfo{person}{Luo Si}.}
  \bibinfo{year}{2017}\natexlab{}.
\newblock \showarticletitle{Cascade Ranking for Operational E-commerce Search}.
  In \bibinfo{booktitle}{\emph{Proceedings of the 23rd ACM SIGKDD International
  Conference on Knowledge Discovery and Data Mining}}.
  \bibinfo{publisher}{ACM}, \bibinfo{address}{New York, NY, USA},
  \bibinfo{pages}{1557--1565}.
\newblock
\showISBNx{978-1-4503-4887-4}


\bibitem[\protect\citeauthoryear{Mnih, Badia, Mirza, Graves, Harley, Lillicrap,
  Silver, and Kavukcuoglu}{Mnih et~al\mbox{.}}{2016}]%
        {mnih2016asynchronous}
\bibfield{author}{\bibinfo{person}{Volodymyr Mnih},
  \bibinfo{person}{Adri{\`a}~Puigdom{\`e}nech Badia}, \bibinfo{person}{Mehdi
  Mirza}, \bibinfo{person}{Alex Graves}, \bibinfo{person}{Tim Harley},
  \bibinfo{person}{Timothy~P Lillicrap}, \bibinfo{person}{David Silver}, {and}
  \bibinfo{person}{Koray Kavukcuoglu}.} \bibinfo{year}{2016}\natexlab{}.
\newblock \showarticletitle{Asynchronous methods for deep reinforcement
  learning}. In \bibinfo{booktitle}{\emph{Proceedings of the 33rd International
  Conference on International Conference on Machine Learning}},
  Vol.~\bibinfo{volume}{48}. JMLR. org, \bibinfo{pages}{1928--1937}.
\newblock


\bibitem[\protect\citeauthoryear{Natarajan}{Natarajan}{1995}]%
        {natarajan1995sparse}
\bibfield{author}{\bibinfo{person}{Balas~Kausik Natarajan}.}
  \bibinfo{year}{1995}\natexlab{}.
\newblock \showarticletitle{Sparse approximate solutions to linear systems}.
\newblock \bibinfo{journal}{\emph{SIAM journal on computing}}
  \bibinfo{volume}{24}, \bibinfo{number}{2} (\bibinfo{year}{1995}),
  \bibinfo{pages}{227--234}.
\newblock


\bibitem[\protect\citeauthoryear{Ng, Harada, and Russell}{Ng
  et~al\mbox{.}}{1999}]%
        {ng1999policy}
\bibfield{author}{\bibinfo{person}{Andrew~Y. Ng}, \bibinfo{person}{Daishi
  Harada}, {and} \bibinfo{person}{Stuart~J. Russell}.}
  \bibinfo{year}{1999}\natexlab{}.
\newblock \showarticletitle{Policy Invariance Under Reward Transformations:
  Theory and Application to Reward Shaping}. In
  \bibinfo{booktitle}{\emph{Proceedings of the Sixteenth International
  Conference on Machine Learning}}. \bibinfo{publisher}{Morgan Kaufmann
  Publishers Inc.}, \bibinfo{address}{San Francisco, CA, USA},
  \bibinfo{pages}{278--287}.
\newblock
\showISBNx{1-55860-612-2}


\bibitem[\protect\citeauthoryear{Pedregosa, Varoquaux, Gramfort, Michel,
  Thirion, Grisel, Blondel, Prettenhofer, Weiss, Dubourg, Vanderplas, Passos,
  Cournapeau, Brucher, Perrot, and Duchesnay}{Pedregosa et~al\mbox{.}}{2011}]%
        {scikit-learn}
\bibfield{author}{\bibinfo{person}{F. Pedregosa}, \bibinfo{person}{G.
  Varoquaux}, \bibinfo{person}{A. Gramfort}, \bibinfo{person}{V. Michel},
  \bibinfo{person}{B. Thirion}, \bibinfo{person}{O. Grisel},
  \bibinfo{person}{M. Blondel}, \bibinfo{person}{P. Prettenhofer},
  \bibinfo{person}{R. Weiss}, \bibinfo{person}{V. Dubourg}, \bibinfo{person}{J.
  Vanderplas}, \bibinfo{person}{A. Passos}, \bibinfo{person}{D. Cournapeau},
  \bibinfo{person}{M. Brucher}, \bibinfo{person}{M. Perrot}, {and}
  \bibinfo{person}{E. Duchesnay}.} \bibinfo{year}{2011}\natexlab{}.
\newblock \showarticletitle{Scikit-learn: Machine Learning in {P}ython}.
\newblock \bibinfo{journal}{\emph{Journal of Machine Learning Research}}
  \bibinfo{volume}{12} (\bibinfo{year}{2011}), \bibinfo{pages}{2825--2830}.
\newblock


\bibitem[\protect\citeauthoryear{Peters and Schaal}{Peters and Schaal}{2008}]%
        {peters2008natural}
\bibfield{author}{\bibinfo{person}{Jan Peters} {and} \bibinfo{person}{Stefan
  Schaal}.} \bibinfo{year}{2008}\natexlab{}.
\newblock \showarticletitle{Natural actor-critic}.
\newblock \bibinfo{journal}{\emph{Neurocomputing}} \bibinfo{volume}{71},
  \bibinfo{number}{7} (\bibinfo{year}{2008}), \bibinfo{pages}{1180--1190}.
\newblock


\bibitem[\protect\citeauthoryear{Schneiderman}{Schneiderman}{2004}]%
        {schneiderman2004feature}
\bibfield{author}{\bibinfo{person}{Henry Schneiderman}.}
  \bibinfo{year}{2004}\natexlab{}.
\newblock \showarticletitle{Feature-centric evaluation for efficient cascaded
  object detection}. In \bibinfo{booktitle}{\emph{Computer Vision and Pattern
  Recognition, 2004. CVPR 2004. Proceedings of the 2004 IEEE Computer Society
  Conference on}}, Vol.~\bibinfo{volume}{2}. IEEE, \bibinfo{pages}{II--II}.
\newblock


\bibitem[\protect\citeauthoryear{Sutton and Barto}{Sutton and Barto}{1998}]%
        {Sutton}
\bibfield{author}{\bibinfo{person}{Richard~S. Sutton} {and}
  \bibinfo{person}{Andrew~G. Barto}.} \bibinfo{year}{1998}\natexlab{}.
\newblock \bibinfo{booktitle}{\emph{Introduction to {R}einforcement {L}earning}
  (\bibinfo{edition}{1st} ed.)}.
\newblock \bibinfo{publisher}{MIT Press}, \bibinfo{address}{Cambridge, MA,
  USA}.
\newblock


\bibitem[\protect\citeauthoryear{Vinyals, Fortunato, and Jaitly}{Vinyals
  et~al\mbox{.}}{2015}]%
        {vinyals2015pointer}
\bibfield{author}{\bibinfo{person}{Oriol Vinyals}, \bibinfo{person}{Meire
  Fortunato}, {and} \bibinfo{person}{Navdeep Jaitly}.}
  \bibinfo{year}{2015}\natexlab{}.
\newblock \showarticletitle{Pointer Networks}.
\newblock In \bibinfo{booktitle}{\emph{Advances in Neural Information
  Processing Systems 28}}, \bibfield{editor}{\bibinfo{person}{C.~Cortes},
  \bibinfo{person}{N.~D. Lawrence}, \bibinfo{person}{D.~D. Lee},
  \bibinfo{person}{M.~Sugiyama}, {and} \bibinfo{person}{R.~Garnett}} (Eds.).
  \bibinfo{publisher}{Curran Associates, Inc.}, \bibinfo{pages}{2692--2700}.
\newblock


\bibitem[\protect\citeauthoryear{Viola and Jones}{Viola and Jones}{2003}]%
        {Viola2003Rapid}
\bibfield{author}{\bibinfo{person}{Paul Viola} {and} \bibinfo{person}{Michael
  Jones}.} \bibinfo{year}{2003}\natexlab{}.
\newblock \showarticletitle{Rapid Object Detection using a Boosted Cascade of
  Simple Features}. In \bibinfo{booktitle}{\emph{Proceedings of the 2001 IEEE
  Computer Society Conference on Computer Vision and Pattern Recognition}}.
  \bibinfo{pages}{I--511--I--518 vol.1}.
\newblock


\bibitem[\protect\citeauthoryear{Wang, Lin, and Metzler}{Wang
  et~al\mbox{.}}{2010a}]%
        {Wang:2010:LER:1835449.1835475}
\bibfield{author}{\bibinfo{person}{Lidan Wang}, \bibinfo{person}{Jimmy Lin},
  {and} \bibinfo{person}{Donald Metzler}.} \bibinfo{year}{2010}\natexlab{a}.
\newblock \showarticletitle{Learning to Efficiently Rank}. In
  \bibinfo{booktitle}{\emph{Proceedings of the 33rd International ACM SIGIR
  Conference on Research and Development in Information Retrieval}}.
  \bibinfo{publisher}{ACM}, \bibinfo{address}{New York, NY, USA},
  \bibinfo{pages}{138--145}.
\newblock
\showISBNx{978-1-4503-0153-4}


\bibitem[\protect\citeauthoryear{Wang, Metzler, and Lin}{Wang
  et~al\mbox{.}}{2010b}]%
        {Wang:2010:RUT:1871437.1871452}
\bibfield{author}{\bibinfo{person}{Lidan Wang}, \bibinfo{person}{Donald
  Metzler}, {and} \bibinfo{person}{Jimmy Lin}.}
  \bibinfo{year}{2010}\natexlab{b}.
\newblock \showarticletitle{Ranking Under Temporal Constraints}. In
  \bibinfo{booktitle}{\emph{Proceedings of the 19th ACM International
  Conference on Information and Knowledge Management}}.
  \bibinfo{publisher}{ACM}, \bibinfo{address}{New York, NY, USA},
  \bibinfo{pages}{79--88}.
\newblock
\showISBNx{978-1-4503-0099-5}


\bibitem[\protect\citeauthoryear{Williams}{Williams}{1992}]%
        {williams1992simple}
\bibfield{author}{\bibinfo{person}{Ronald~J Williams}.}
  \bibinfo{year}{1992}\natexlab{}.
\newblock \showarticletitle{Simple statistical gradient-following algorithms
  for connectionist reinforcement learning}.
\newblock \bibinfo{journal}{\emph{Machine learning}} \bibinfo{volume}{8},
  \bibinfo{number}{3-4} (\bibinfo{year}{1992}), \bibinfo{pages}{229--256}.
\newblock


\bibitem[\protect\citeauthoryear{Xu and Li}{Xu and Li}{2007}]%
        {Xu2007AdaRank}
\bibfield{author}{\bibinfo{person}{Jun Xu} {and} \bibinfo{person}{Hang Li}.}
  \bibinfo{year}{2007}\natexlab{}.
\newblock \showarticletitle{AdaRank: A Boosting Algorithm for Information
  Retrieval}. In \bibinfo{booktitle}{\emph{Proceedings of the 30th Annual
  International ACM SIGIR Conference on Research and Development in Information
  Retrieval}}. \bibinfo{publisher}{ACM}, \bibinfo{address}{New York, NY, USA},
  \bibinfo{pages}{391--398}.
\newblock
\showISBNx{978-1-59593-597-7}


\bibitem[\protect\citeauthoryear{Zheng, Chen, Sun, and Zha}{Zheng
  et~al\mbox{.}}{2007}]%
        {Zheng2007A}
\bibfield{author}{\bibinfo{person}{Zhaohui Zheng}, \bibinfo{person}{Keke Chen},
  \bibinfo{person}{Gordon Sun}, {and} \bibinfo{person}{Hongyuan Zha}.}
  \bibinfo{year}{2007}\natexlab{}.
\newblock \showarticletitle{A Regression Framework for Learning Ranking
  Functions Using Relative Relevance Judgments}. In
  \bibinfo{booktitle}{\emph{Proceedings of the 30th Annual International ACM
  SIGIR Conference on Research and Development in Information Retrieval}}.
  \bibinfo{publisher}{ACM}, \bibinfo{address}{New York, NY, USA},
  \bibinfo{pages}{287--294}.
\newblock
\showISBNx{978-1-59593-597-7}


\bibitem[\protect\citeauthoryear{Zhou}{Zhou}{2012}]%
        {zhou2012ensemble}
\bibfield{author}{\bibinfo{person}{Zhi-Hua Zhou}.}
  \bibinfo{year}{2012}\natexlab{}.
\newblock \bibinfo{booktitle}{\emph{Ensemble methods: foundations and
  algorithms}}.
\newblock \bibinfo{publisher}{CRC press}.
\newblock


\end{thebibliography}

\end{document}